# Spatiotemporal Gaussian representation-based dynamic reconstruction and motion estimation framework for time-resolved volumetric MR imaging (DREME-GSMR)

**Running Title:**
**Time-resolved MRI Reconstruction Via Spatiotemporal Gaussians**

Jiacheng Xie
Hua-Chieh Shao
Can Wu
Ricardo Otazo
Jie Deng
Mu-Han Lin
Tsuicheng Chiu
Jacob Buatti
Viktor Iakovenko
You Zhang

*The Advanced Imaging and Informatics for Radiation Therapy (AIRT) Laboratory*
*The Medical Artificial Intelligence and Automation (MAIA) Laboratory*
*Department of Radiation Oncology, University of Texas Southwestern Medical Center, Dallas, TX 75390, USA*

Corresponding address:

You Zhang
Department of Radiation Oncology
University of Texas Southwestern Medical Center
2280 Inwood Road
Dallas, TX 75390
Email: You.Zhang@UTSouthwestern.edu
Tel: (214) 645-2699

**Abstract**

*Objective.* Time-resolved volumetric MR imaging that reconstructs a 3D MRI within sub-seconds to resolve deformable motion is essential for motion-adaptive radiotherapy. Representing patient anatomy and associated motion fields as 3D Gaussians, we developed a spatiotemporal Gaussian representation-based framework (DREME-GSMR), which enables time-resolved dynamic MRI reconstruction from a pre-treatment 3D MR scan without any prior anatomical/motion model. *Approach.* DREME-GSMR represents a reference MRI volume and a corresponding low-rank motion model (as motion-basis components) using 3D Gaussians, and incorporates a dual-path MLP/CNN motion encoder to estimate temporal motion coefficients of the motion model from raw k-space-derived signals. Furthermore, using the solved motion model, DREME-GSMR can infer motion coefficients directly from new online k-space data, allowing subsequent intra-treatment volumetric MR imaging and motion tracking (real-time imaging). A motion-augmentation strategy is further introduced to improve robustness to unseen motion patterns during real-time imaging. *Main results.* DREME-GSMR was evaluated on the XCAT digital phantom, a physical motion phantom, and MR-LINAC datasets acquired from 6 healthy volunteers and 20 patients (with independent sequential scans for cross-evaluation). DREME-GSMR produced time-resolved MRIs at a ~400 ms temporal resolution, with an inference time of ~10 ms/volume. In XCAT experiments, DREME-GSMR achieved mean(s.d.) SSIM, tumor center-of-mass error (COME), and DICE similarity coefficient (DSC) of 0.92(0.01)/0.91(0.02), 0.50(0.15)/0.65(0.19) mm, and 0.92(0.02)/0.92(0.03) for dynamic reconstruction/real-time imaging. For the physical phantom, the mean target COME was 1.19(0.94)/1.40(1.15) mm for dynamic/real-time imaging, while for volunteers and patients, the mean liver COME for real-time imaging was 1.31(0.82) and 0.96(0.64) mm, respectively. *Significance.* Evaluated on digital simulations and MR-LINAC measurements, DREME-GSMR enables sub-second 3D dynamic MRI reconstruction and real-time volumetric motion tracking from stack-of-stars k-space data, supporting intra-treatment motion-adaptive MR-guided radiotherapy.

## 1. Introduction

Magnetic resonance imaging (MRI) provides superior soft tissue contrast for anatomical visualization without ionizing radiation. It has been widely integrated into clinical workflows to enable MRI-guided radiotherapy (Corradini *et al.*, 2019; Hall *et al.*, 2019; Owrangi *et al.*, 2018). For anatomical sites with respiratory motion (Bertholet *et al.*, 2016; Seppenwoolde *et al.*, 2002), especially the thoracic and abdominal regions, four-dimensional MRI (4D-MRI) was developed for motion management (Stemkens *et al.*, 2018). For a typical 4D-MRI acquisition, a respiratory motion surrogate signal sorts continuously acquired k-space data into predefined respiratory bins to reconstruct volumetric MRIs from each bin, yielding a motion-resolved sequence of three-dimensional (3D) images for a representative breathing cycle (Menchón-Lara *et al.*, 2019; Rajiah *et al.*, 2023; Stemkens *et al.*, 2018). By analyzing patient-specific 4D-MRIs, clinicians can tailor motion-mitigation strategies, such as breath-hold techniques or respiratory gating, to reduce target localization uncertainty (Ball *et al.*, 2022; Paganelli *et al.*, 2018). However, 4D-MRI has inherent limitations. Its respiratory sorting process assumes periodic and reproducible breathing, an assumption often invalidated by irregular motion patterns (e.g., amplitude variations, frequency shifts, baseline drifts), leading to mis-sorted data and degraded image quality, and failing to capture irregular motion patterns that may affect motion management decisions.

To overcome these shortcomings, *time-resolved dynamic* volumetric MRI retrospectively reconstructs a sequence of 3D images at a high temporal resolution (Nayak, 2019), without using motion surrogates for sorting. Each frame is reconstructed from a highly undersampled subset of k-space data (e.g., tens of radial spokes), effectively freezing the anatomy and eliminating intra-scan motion. Recent single-scan-based frameworks jointly reconstruct all time-frames by exploiting global spatiotemporal correlations, thereby addressing the extreme undersampling issue. For example, MR-MOTUS (Huttinga *et al.*, 2021a; Huttinga *et al.*, 2020) estimates non-rigid 3D motion directly from undersampled k-space and reconstructs dynamic images by deforming a reference MRI. Extreme MRI (Ong *et al.*, 2020) uses multiscale low-rank factorization to model continuous non-gated volumetric dynamics. STINR-MR jointly reconstructs a reference image and motion fields using spatial and temporal implicit neural representations with a PCA-based motion model (Shao *et al.*, 2024a). Although dynamic volumetric MRI provides richer motion information than conventional sorting-based 4D-MRI, existing reconstruction methods are not designed for beam-on applications because they typically require access to the full acquisition for reconstruction. To enable real-time treatment verification and adaptation (Keall *et al.*, 2025; McNair and Buijs, 2019; Keall *et al.*, 2019), *time-resolved real-time* MR imaging is needed to capture anatomical changes with a sub-second latency (Bertholet *et al.*, 2019; Nayak *et al.*, 2022; Lombardo *et al.*, 2024). For respiratory motion management, the recommended end-to-end latency, including both image acquisition and reconstruction, is within 500 milliseconds (ms) (Keall *et al.*, 2021a). This stringent requirement largely precludes the use of time-consuming dynamic MRI reconstruction algorithms on real-time data and necessitates highly efficient reconstruction and motion tracking strategies.

Existing methods for real-time volumetric MR imaging can be broadly categorized into DL-based registration approaches, offline-online motion-model methods, and dual-task learning methods. DL-based registration methods, such as TEMPEST, KS-RegNet, and the method by Wei et al., estimate 3D motion by registering prior MR images to undersampled dynamic MR data (Wei *et al.*, 2023; Shao *et al.*, 2022; Terpstra *et al.*, 2021). However, patient-specific prior images or motion models may bias the motion estimation, when the prior anatomy/contrast and/or motion model become outdated. High-quality MR datasets are also required for training these DL models, limiting generalizability to out-of-distribution scenarios. Another line of work adopts an offline-online strategy, such as real-time MR-MOTUS and MRSIGMA, which first construct patient-specific anatomy and motion models from offline 4D-MRI, and then use these models for rapid online motion estimation (Huttinga *et al.*, 2022; Feng *et al.*, 2020). Although efficient during inference, their reliance on predefined motion states limits their ability to fully represent irregular motion.

More recently, a Dynamic Reconstruction and Motion Estimation for MRI (DREME-MR) framework was proposed (Shao *et al.*, 2025b). Based on STINR-MR (Shao et al., 2024a) and the X-ray-based DREME framework (Shao *et al.*, 2025a), DREME-MR introduced a dual-task, one-shot learning framework that jointly performs dynamic MRI reconstruction and real-time motion estimation. Validated on digital phantoms and a human subject, DREME-MR achieved 3 mm spatial and 100-150 ms temporal resolutions for cardiorespiratory motion without requiring patient-specific priors or respiratory binning. However, its major drawback is training time (~ 70 min on an NVIDIA 4090 GPU), which may limit online clinical deployment. In addition, its real-time imaging capability for the human subject was evaluated on a single MRI scan, by using 75% of the acquired data for training and the remaining 25% for testing. As a result, the reported performance primarily reflects the model's ability to generalize within the same acquisition (intra-scan) rather than a truly inter-scan real-time setting, where training and inference data should be collected sequentially and independently. Consequently, its evaluation remains relatively narrow. Furthermore, the MRI data used in DREME-MR were acquired with a 3D golden-mean Koosh-ball radial trajectory, which has mainly been reported in research or prototype implementations (Feng, 2022; Huttinga *et al.*, 2021b). In contrast, current clinical MR-LINAC workflows more commonly rely on stack-of-stars (SOS) radial sequences for motion-impacted sites like the abdomen (Feng, 2022; Keesman *et al.*, 2024; Ferris *et al.*, 2024; Daly *et al.*, 2024; Paulson *et al.*, 2020). Beyond its clinical prevalence, SOS is also well-suited to time-resolved imaging. By combining radial sampling in the $k_x - k_y$ plane with Cartesian encoding along $k_z$ (Feng, 2022), and rotating successive stacks (a radial stack refers to all spokes at different $k_z$ acquired at one in-plane angle) by a golden angle, SOS sampling inherits the motion robustness and flexible retrospective grouping of radial sampling while preserving the efficiency of Cartesian encoding along $k_z$. Importantly, SOS supports self-navigation, from k-space-center self-gating signals (Li *et al.*, 2023; Feng, 2023a; Feng *et al.*, 2020; Zhang *et al.*, 2019; Feng *et al.*, 2016) to image-domain projections (Chen *et al.*, 2024b; Chen *et al.*, 2024a; Feng, 2023b), making it a versatile acquisition strategy for time-resolved MRI .

From a representation perspective, DREME-MR relies on an implicit neural representation (INR)-based reference MRI representation and B-spline-parameterized motion-basis components. An INR encodes an image as a continuous function rather than a conventional voxelized representation, typically through multilayer perceptrons (MLPs), to map spatial coordinates to intensity values (Mildenhall *et al.*, 2021). Such a representation is compact and can be fit through self-supervised optimization to reconstruct images directly from sparsely sampled measurements without large-scale training data (Shen *et al.*, 2022). This continuous encoding, however, requires time-consuming coordinate querying and interpolation operations during patient-specific optimization. It motivates the use of more explicit and computationally efficient volumetric representations. Recently, 3D Gaussian representation (Fei *et al.*, 2024) has been applied in medical image reconstruction for X-ray-based CT (Zha *et al.*, 2024) and MRI reconstruction (Peng *et al.*, 2025a; Wu *et al.*, 2024). In this approach, anatomy is represented as a collection of 3D Gaussian functions instead of dense voxel grids or neural networks, preserving the continuous volumetric properties of anatomy while avoiding unnecessary computations in empty spaces. It has demonstrated advantages over INR-based methods in both reconstruction speed and accuracy for CT (Li *et al.*, 2025; Lin *et al.*, 2024; Zha *et al.*, 2024) and MRI (Wu *et al.*, 2024). Gaussian-based methods employ explicit basis functions, Gaussian primitives, to directly represent the target volume, offering a greater capacity to preserve high-frequency anatomical details such as bony edges. For example, PMF-STGR (Xie *et al.*, 2025) employed 3D Gaussians to represent both the reference image and motion bases in dynamic CBCT reconstruction, resulting in an approximately 50% reduction in reconstruction time and memory consumption compared with INR-based formulations. These advantages have also motivated the extension of Gaussian-based representations from static to 4D-MRI reconstruction. In this direction, MoRe-3DGSMR (Peng *et al.*, 2025b) reconstructs 4D-MRIs by first reconstructing a reference image of a motion state using a Gaussian representation and then warping it by deformation vector fields (DVFs) resolved from a convolutional neural network to generate volumes of the remaining motion states. While achieving good reconstruction accuracy, it falls within the conventional 4D-MRI framework and requires respiratory sorting.

In this study, we propose a new framework, DREME-GSMR, to introduce the strong representation power of 3D Gaussians into the time-resolved MRI reconstruction pipeline. DREME-GSMR jointly represents a reference-frame MRI and corresponding coarse-to-fine motion-basis components (MBCs) with 3D Gaussians, while a dual-path MLP/CNN motion encoder estimates the temporal coefficients of MBCs from k-space-derived signals. This design enables two coupled tasks within one patient-specific training process: dynamic MRI reconstruction from a pre-treatment scan and subsequent real-time volumetric motion estimation from newly acquired online k-space data. The main contributions are threefold. First, DREME-GSMR achieves high-quality reconstruction and accurate motion estimation while reducing training time by 30% compared with INR-based DREME-MR. Second, its MLP/CNN dual-path motion encoder incorporates, to our best knowledge, the first motion-augmentation strategy designed for real-time MRI, enabling robust real-time motion estimation beyond the training distribution. Third, DREME-GSMR was extensively evaluated on digital phantom simulations, physical phantom measurements, and clinical datasets, including 6 healthy volunteers and 20 patients. All measured data were acquired using a clinically available golden-angle SOS sequence on an MR-LINAC system. Importantly, each clinical case contains two sequential scans, allowing training and testing on independent acquisitions for cross-evaluation and providing a more realistic assessment of real-time imaging performance.

## 2. Methods

### *2.1 Dynamic and real-time volumetric MR imaging overview*

For DREME-GSMR, dynamic MRI reconstruction aims to reconstruct a sequence of dynamic images to visualize the 3D anatomical motion from a pre-treatment scan for treatment guidance. Instead of independently reconstructing each highly undersampled 3D frame, the dynamic sequence is represented by a reference-frame MRI $\boldsymbol{I}_{\text{ref}}(\boldsymbol{x})$ and a set of time-dependent DVFs $\boldsymbol{d}(\boldsymbol{x},t)$ describing intra-scan motion relative to this reference. Each frame corresponds to a short acquisition window in which anatomical motion is assumed to be negligible. Therefore, by deforming the reference-frame MRI $\boldsymbol{I}_{\text{ref}}(\boldsymbol{x})$ with DVFs $\boldsymbol{d}(\boldsymbol{x},t)$ that represent the anatomical motion at each time frame $t$, the dynamic MRI sequence $\boldsymbol{I}(\boldsymbol{x},t)$ can be obtained as:

$$\boldsymbol{I}(\boldsymbol{x},t) = \boldsymbol{I}_{\text{ref}}\big(\boldsymbol{x} + \boldsymbol{d}(\boldsymbol{x},t)\big), \tag{1}$$

where $\boldsymbol{x} \in \mathbb{R}^3$ denotes the voxel coordinates of the reconstruction volume. For further dimension reduction to address the ill-posed problem, $\boldsymbol{d}(\boldsymbol{x},t)$ can be approximated by a low-rank function (Li *et al.*, 2011; Stemkens *et al.*, 2016; Zhang *et al.*, 2007) via a summation of products of spatial $\boldsymbol{e}_i(\boldsymbol{x})$ and temporal $\boldsymbol{w}_i(t)$ components:

$$\boldsymbol{d}(\boldsymbol{x},t) = \sum_{i=1}^{3} \boldsymbol{w}_i(t) \times \boldsymbol{e}_i(\boldsymbol{x}). \tag{2}$$

The spatial component $\boldsymbol{e}_i(\boldsymbol{x})$ serves as a basis set that spans the motion space, and all motion states of the dynamic sequence can be described by scaling the spatial components via the temporal components $\boldsymbol{w}_i(t)$. Accordingly, $\boldsymbol{e}_i(\boldsymbol{x})$ and $\boldsymbol{w}_i(t)$ are called motion basis components (MBCs) and MBC scores, respectively. Thus, dynamic MR imaging is equivalent to reconstructing a reference-frame MRI $\boldsymbol{I}_{\text{ref}}$, while determining time-varying MBC scores $\boldsymbol{w}_i(t)$ of MBCs $\boldsymbol{e}_i(\boldsymbol{x})$ that capture the underlying anatomical motion. Following prior studies (Li *et al.*, 2011; Shao *et al.*, 2024a; Shao *et al.*, 2025b), three MBCs (i.e. $i = 1, 2, 3$) for each Cartesian direction were used to model breathing motion.

With the above strategies, the dynamic reconstruction is formulated as an optimization problem:

$$\hat{\boldsymbol{I}}(\boldsymbol{x},t) = \arg\min_{\boldsymbol{I}_{\text{ref}}(x),\, \boldsymbol{w}_i(t),\, \boldsymbol{e}_i(x)} \mathbb{E}\,[||\mathcal{F}[\boldsymbol{I}(\boldsymbol{x},t)] - s(\boldsymbol{k},t)||_1] + \lambda R[\boldsymbol{I}(\boldsymbol{x},t)], \tag{3}$$

where $\boldsymbol{s}(\boldsymbol{k},t)$ denotes the acquired k-space signal at the frame $t$, $\mathcal{F}$ denotes the operator combining the coil sensitivity encoding and non-uniform fast Fourier transform (NUFFT), $||\cdot||_1$ denotes the L1 norm, $\mathbb{E}$ denotes the expectation value, and $\lambda$ is the weighting factor of the regularization term $R$. The first term of Eq. 3 enforces the k-space data consistency by matching the reconstructed k-space data with the acquired k-space signals $\boldsymbol{s}(\boldsymbol{k},t)$. The second term regularizes the optimization process to mitigate the undersampled reconstruction challenge (see Sec. 2.3.3 for details). After the dynamic reconstruction, the learned reference-frame $\boldsymbol{I}_{\mathrm{ref}}(\boldsymbol{x})$ and the corresponding motion model (i.e., $\boldsymbol{e}_i(\boldsymbol{x})$ and $\boldsymbol{w}_i(t)$) would allow DREME-GSMR to infer the MBC scores $\boldsymbol{w}_i(t)$ from future limited-sampled k-space acquisitions $\boldsymbol{s}(\boldsymbol{k},t)$ for real-time imaging (Sec. 2.3.1).

### *2.2 Gaussian representation*

We represent the target objects with a set of learnable 3D Gaussian kernels $\mathbb{G}^3 = \{G_i^3\}_{i=1,\ldots,M}$ such that each kernel $G_i^3$ defines a local Gaussian-shaped density field:

$$G_i^3(\boldsymbol{x} \mid \rho_i, \boldsymbol{p}_i, \boldsymbol{\Sigma}_i) = \rho_i \cdot \exp\left(-\frac{1}{2}(\boldsymbol{x}-\boldsymbol{p}_i)^\top \boldsymbol{\Sigma}_i^{-1}(\boldsymbol{x}-\boldsymbol{p}_i)\right), \tag{4}$$

where $\boldsymbol{x} \in \mathbb{R}^3$ denotes the voxel coordinates of the represented volume. $\rho_i$ , $\boldsymbol{p}_i \in \mathbb{R}^3$ and $\boldsymbol{\Sigma}_i \in \mathbb{R}^{3\times3}$ are learnable parameters representing the central density, the center location, and the shape size and orientation of each Gaussian point, respectively. The overall density $\sigma(\boldsymbol{x})$ at $\boldsymbol{x} \in \mathbb{R}^3$ can be obtained by summing the densities of kernels, which is the voxelization operation for Gaussians:

$$\sigma(\boldsymbol{x}) = \sum_{i=1}^{M} G_i^3(\boldsymbol{x} \mid \rho_i, \boldsymbol{p}_i, \boldsymbol{\Sigma}_i). \tag{5}$$

The above formulations can only model real-valued objects. For complex-valued target objects like MR images, we represent them with a set of learnable complex-valued 3D Gaussian kernels, where each kernel uses two independent density coefficients ${\rho_i}^{Re}$ and ${\rho_i}^{Im}$ to model the real and imaginary components, respectively:

$$\tilde{G}_i^3(\boldsymbol{x} \mid {\rho_i}^{Re}, {\rho_i}^{Im}, \boldsymbol{p}_i, \boldsymbol{\Sigma}_i) = ({\rho_i}^{Re} + j{\rho_i}^{Im}) \cdot \exp\left(-\frac{1}{2}(\boldsymbol{x}-\boldsymbol{p}_i)^\top \boldsymbol{\Sigma}_i^{-1}(\boldsymbol{x}-\boldsymbol{p}_i)\right). \tag{6}$$

The overall complex-valued density can be obtained by:

$$\tilde{\sigma}(\boldsymbol{x}) = \sum_{i=1}^{M} \tilde{G}_i^3(\boldsymbol{x} \mid {\rho_i}^{Re}, {\rho_i}^{Im}, \boldsymbol{p}_i, \boldsymbol{\Sigma}_i), \tag{7}$$

which can be decomposed into real and imaginary parts as $\tilde{\sigma}(\boldsymbol{x}) = \sigma^{Re}(\boldsymbol{x}) + j\sigma^{Im}(\boldsymbol{x})$. In DREME-GSMR, both the reference-frame MRI and the MBCs are represented using Gaussians with signed (negative-permitting) density coefficients, enabled by a modified CUDA-based voxelizer adapted from Zha et al. (Zha *et al.*, 2024). For the reference-frame MRI $\boldsymbol{I}_{\mathrm{ref}}$, we use complex-valued 3D Gaussians, in contrast to the real-valued MBC Gaussians.

### *2.3 The DREME-GSMR method*

#### *2.3.1 Overview of DREME-GSMR and dual-task learning*

**Figure 1** illustrates the workflow of DREME-GSMR, which comprises three primary components: reference-frame MRI Gaussians, MBC Gaussians, and a dual-path MLP/CNN-based motion encoder, which represent the reference-frame volume $\boldsymbol{I}_{\mathrm{ref}}$, MBCs $\boldsymbol{e}_i(\boldsymbol{x})$, and MBC scores $\boldsymbol{w}_i(t)$, respectively. DREME-GSMR is trained in a 'one-shot' fashion, using only the k-space acquisition $\boldsymbol{s}(\boldsymbol{k},t)$ of a pre-treatment MR scan to generate the time-resolved dynamic sequence of volumetric MR images $\boldsymbol{I}(\boldsymbol{x},t)$. The dual-task learning is guided by data consistency and regularization losses as defined in Eq. 3. After the dual-task learning, DREME-GSMR yields the up-to-date reference-frame MRI $\boldsymbol{I}_{\mathrm{ref}}$, a patient-specific motion model (i.e., MBCs $\boldsymbol{e}_i(\boldsymbol{x})$ and k-space signal-dependent MBC scores $\boldsymbol{w}_i(t)$), and a motion encoder that can infer $\boldsymbol{w}_i$ based on k-space signals. For real-time imaging, DREME-GSMR takes newly acquired online MR signals to estimate the real-time DVF $\boldsymbol{d}(\boldsymbol{x},t)$, to apply to the reference-frame MRI to derive the real-time MRI. Additionally, tracking targets (e.g. tumor) can be contoured in the reference-frame MRI $\boldsymbol{I}_{\mathrm{ref}}$ to obtain a target mask, and then the real-time DVF $\boldsymbol{d}(\boldsymbol{x}, t)$ is applied to the target mask to achieve markerless target localization.

The reference-frame MRI $\boldsymbol{I}_{\mathrm{ref}}$ is encoded by a single set of complex-valued 3D Gaussians (Eq. 6), reflecting the complex-valued nature of raw MR data. The MBCs $\boldsymbol{e}_i(\boldsymbol{x})$ of the motion model are represented using MBC Gaussians at three spatial levels along each Cartesian direction (x, y, z), resulting in a total of nine MBC volumes, with point counts increasing across levels to capture coarse-to-fine motion. Both are voxelized through a differentiable CUDA voxelizer (Eq. 5 and Eq. 7) and optimized using gradients from the data-consistency and regularization losses defined in Eq. 3, while the dual-path encoder and its inputs are detailed in Sec. 2.3.2.

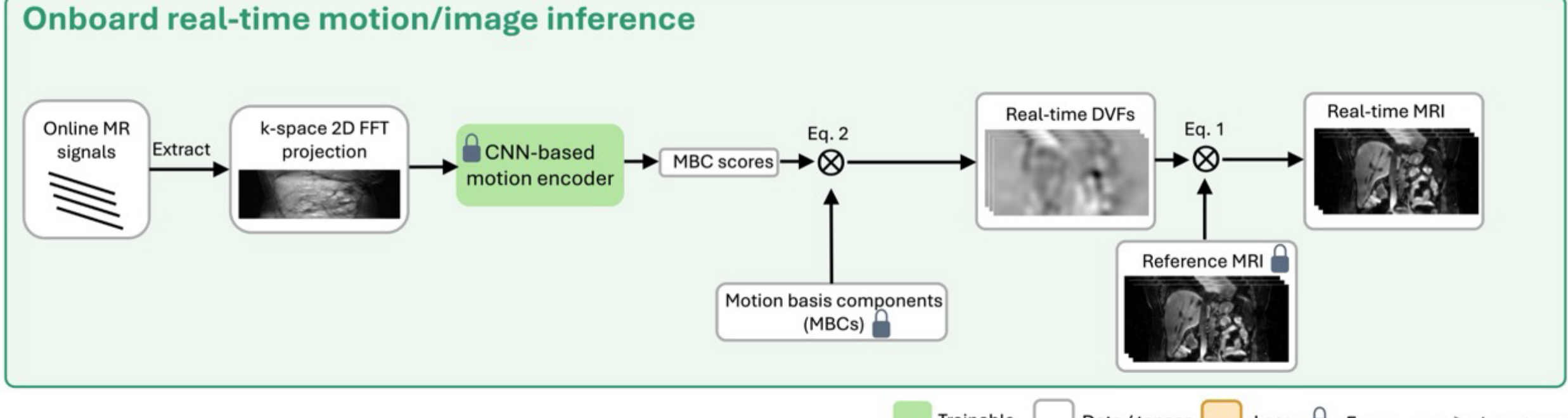

**Figure 1. Overview of DREME-GSMR**. In the *training stage*, DREME-GSMR simultaneously reconstructs a dynamic MRI sequence and trains a dual-path motion encoder for real-time motion estimation, all based on a pre-treatment MR scan. After training, a motion-compensated reference-frame MRI is solved along with the MBCs and the motion encoder. In the *onboard real-time imaging stage*, the reconstructed reference-frame MRI, respiratory MBCs, and the motion encoder are used to derive real-time MRI, using real-time k-space signals. This figure presents a general overview of the DREME-GSMR framework, and the detailed training and inference setups can be found in the following **Fig. 2**.

*2.3.2 Dual-path motion encoder*

For the design of the dual-path MLP/CNN motion encoder, the MLP path captures global bulk motion based on the k-space center, while the CNN path models finer dynamics from the k-space data and enables motion augmentation for robust real-time motion estimation. For the CNN-based motion encoder, prior work has shown that training with motion-augmented simulated data enables extrapolation beyond the motion range represented in the original dynamic sequence (Shao *et al.*, 2025a). This capability is particularly important for real-time imaging, where motion encountered during testing may differ substantially from that observed during training.

The MLP-based motion encoder was designed to directly take the k-space signals $\boldsymbol{s}(\boldsymbol{k}, t)$ as input to estimate the MBC scores $\boldsymbol{w}_i(t)$. Because each coil in a phased-array acquisition has a distinct spatial sensitivity profile, the corresponding k-space signals inherently encode localized anatomical motion. In the SOS trajectory, every readout crosses the in-plane k-space center, such that motion-sensitive samples can be extracted at the central readout position ($k_x = k_y = 0$) across several near-central $k_z$ ($k_z \sim 0$) partitions. Accordingly, for each stack, the complex-valued samples at the in-plane k-space center of selected central $k_z$ locations ($|k_z| \leq |k_{max}|$) were extracted from all coils, and their real and imaginary components were concatenated to form the input to the MLP encoder. The MLP-based motion encoder has 9 MLP networks, corresponding to the three-level respiratory MBCs $\boldsymbol{e}_i(\boldsymbol{x})$ (with three Cartesian components each). All MLPs share an identical architecture composed of three fully connected layers. The first two layers are followed by rectified linear unit (ReLU) activations, while the final layer remains linear to yield a scalar output corresponding to a specific motion component.

The CNN-based motion encoder was designed to enable image-domain motion estimation. For each SOS stack, a central k-space patch of size $3 \times 6$ was retained, while the remaining k-space samples were zero-filled. A 2D inverse FFT was then applied coil-wise, and the resulting coil images were combined using root-sum-of-squares to generate a single-channel projection image for motion estimation. By restricting the input to the central $3 \times 6$ k-space patch for each stack, the encoder was encouraged to focus on dominant smooth motion information while reducing sensitivity to fine-scale high-frequency image details, thereby avoiding model overfitting, consistent with prior strategies that derive motion surrogates from the central k-space region (Ahmed *et al.*, 2020; Jiang *et al.*, 2018). The CNN encoder comprised six convolutional blocks with kernel sizes of $7 \times 7, 7 \times 7, 5 \times 5, 5 \times 5, 3 \times 3,$ and $3 \times 3$, and channel numbers of 2, 4, 8, 16, 32, and 32, respectively. Each block consisted of a stride-2 convolution, followed by batch normalization and ReLU activation. The final feature maps were globally averaged and passed through a fully connected layer to predict 9 MBC scores, corresponding to the motion-basis coefficients along different spatial directions and basis levels.

*2.3.3 Regularization loss functions*

To ensure physiologically plausible motion modeling, several regularization losses were incorporated during the training of DREME-GSMR. The first is a total variation (TV) loss (Zhang *et al.*, 2017), which mitigates high-frequency noise while preserving anatomical structures and edge integrity:

$$L_{TV} = \frac{1}{N_{voxel}} \sum_i |\nabla \boldsymbol{I}_{\mathrm{ref}}(x_i)|, \tag{8}$$

where $N_{voxel}$ denotes the number of voxels, $i$ denotes the voxel index, and $\nabla$ denotes the gradient operator.

The other loss functions were introduced to regularize the respiratory motion model. Firstly, to mitigate ambiguities in the spatiotemporal decomposition (Eq. 2) of the low-rank motion model, we incorporate a normalization loss to promote the normality of the MBCs $\boldsymbol{e}_i(\boldsymbol{x})$ :

$$L_{MBC} = \frac{1}{9} \sum_{k=x,y,z} \sum_{i=1}^{3} \left( \left| ||e_{i,k}||_2^2 - 1 \right|^2 \right), \tag{9}$$

where $||\cdot||_2$ is the L2 norm. Secondly, a zero-mean regularization is applied to the MBC scores $\boldsymbol{w}_i(t)$:

$$L_{ZMS} = \frac{1}{9} \sum_{k=x,y,z} \sum_{i=1}^{3} \left| \frac{1}{N_t} \sum_t w_{i,k}(t) \right|^2. \tag{10}$$

This loss penalizes baseline offsets in the MBC scores, effectively placing the mean of the estimated respiratory motion around the origin. Consequently, this loss ensures that the multiplicative motion augmentation (see Sec. 2.3.4 for details) scales the motion symmetrically about the origin, producing an evenly distributed range of augmented motion states rather than one biased toward a single direction. Thirdly, a Jacobian regularization term was introduced to suppress anatomically implausible organ motion (Chen *et al.*, 2025; Heyde *et al.*, 2016; Chun and Fessler, 2009) by constraining the local deformation induced by the dynamic DVFs:

$$L_{Jac} = \frac{1}{\sum_t |\Omega_t|} \sum_t \sum_{x \in \Omega_t} \left( det\left( J_{\emptyset_t}(\boldsymbol{x}) \right) - 1 \right)^2, \tag{11}$$

where $J_{\emptyset_t}(\boldsymbol{x}) = \nabla \emptyset_t(\boldsymbol{x})$ represents the Jacobian matrix of the transformation $\emptyset_t(\boldsymbol{x}) = \boldsymbol{x} + \boldsymbol{d}(\boldsymbol{x}, t)$ at point $\boldsymbol{x}$, while $\Omega_t$ denotes the region of interest (ROI) for each frame $t$. An ROI mask was generated from the reference image using intensity thresholding to separate the low-intensity air/background and focus the Jacobian regularization on the anatomy. This form penalizes large deviations of the determinant of the Jacobian matrix from 1, thereby discouraging unrealistic local compression, expansion, and folding in the estimated DVF. Finally, a frequency regularization term was incorporated to suppress non-physiological oscillations in the motion scores inferred by the motion encoder by penalizing abnormal spectral peaks (as demonstrated in **Appendix: Fig. A.4**). In detail, the temporal scores were transformed into the frequency domain, and the average spectral power in predefined frequency bands was compared with that in neighboring baseline bands:

$$L_{Freq} = \frac{1}{N_{peak}} \sum |\bar{P}(C) - \bar{P}(B)|, \tag{12}$$

where $\bar{P}(C)$ and $\bar{P}(B)$ denote the mean spectral power within the target frequency band $C$ and its corresponding neighboring baseline band $B$, respectively, and $N_{peak}$ denotes the number of predefined frequency bands. Each score sample of the motion encoder output was associated with a stack, and each stack spanned $n_{kz}$ spokes with a repetition time (TR) of $t_r$. Therefore, the score sampling rate was $f_s = \frac{1}{t_r n_{kz}}$, yielding a Nyquist frequency of $f_N = \frac{1}{2t_r n_{kz}}$. A predefined scanner-related reference frequency (60.7353 Hz, reflecting acquisition noise), its higher-order harmonics, and their folded aliases were then mapped into this Nyquist range $f_N$ to define the target bands, while adjacent side bands served as local spectral baselines.

*2.3.4 Training strategy*

To initialize the Gaussians, a motion-averaged MRI $\boldsymbol{I}_{\text{avg}}(\boldsymbol{x})$ is reconstructed via inverse NUFFT using coil-compressed k-space data of all radial spokes. For reference-frame MRI Gaussians, $M$ points are sampled uniformly from $\boldsymbol{I}_{\text{avg}}(\boldsymbol{x})$, preserving their corresponding positions and densities. For MBC Gaussians, $M_1$, $M_2$, and $M_3$ points are sampled using grid-based sampling for three spatial levels to represent coarse-to-fine ($M_3 > M_2 > M_1$) motion. With the negative-permitting voxelizer, we used one complex-valued Gaussian cloud to represent the reference-frame MRI, and 9 (3 × 3) real-valued Gaussian clouds to represent the MBCs, one Gaussian cloud for each spatial level and Cartesian direction. As shown in **Fig. 2**, DREME-GSMR follows a progressive four-stage training strategy after initialization to improve learning stability and avoid convergence to local optima, inspired by prior dynamic (Xie *et al.*, 2025; Shao *et al.*, 2024a) and real-time motion estimation (Shao *et al.*, 2025b; Shao *et al.*, 2025a) frameworks.

Stage I (**Fig. 2(a)**) optimizes the reference-frame MRI $\boldsymbol{I}_{ref}$ in two steps: an image-domain L1 fit (Stage I(a)) followed by a k-space data-consistency refinement that suppresses coil-compression artifacts (Stage I(b)), with no anatomical motion considered. In Stage I(a), the loss function is defined in the image domain between the initial motion-averaged MRI $\boldsymbol{I}_{\text{avg}}(\boldsymbol{x})$ and the $\boldsymbol{I}_{\text{ref}}(\boldsymbol{x})$, with $\boldsymbol{I}_{ref}$ obtained via voxelizing the reference-frame Gaussians:

$$L_I^a = \frac{1}{N_{voxel}} \sum_x ||\boldsymbol{I}_{ref}(x) - \boldsymbol{I}_{avg}(x)||_1. \tag{13}$$

Since $\boldsymbol{I}_{\text{avg}}(\boldsymbol{x})$ is reconstructed from coil-compressed k-space data, it exhibits artifacts arising from coil compression. In Stage I(b), the loss function is reformulated as a k-space data consistency loss computed over all acquired k-space samples to suppress the coil compression artifacts, combined with the TV regularization loss (Eq. 8):

$$L_I^b = \mathbb{E}\left[||\mathcal{F}\left[\boldsymbol{I}_{ref}(\boldsymbol{x})\right] - \boldsymbol{s}(\boldsymbol{k})||_1\right] + \lambda_{TV} L_{TV}, \tag{14}$$

where $\lambda_{TV}$ denotes the weighting factor of the TV loss. $\boldsymbol{I}_{ref}$ is converted to k-space via NUFFT for loss calculation.

In Stage II (**Fig. 2(b)**), the reference-frame MRI Gaussians (Eq. 6), the MBC Gaussians (Eq. 4), and the MLP motion encoder (Sec. 2.3.2) are trained simultaneously with the k-space data consistency loss and the regularization losses (Eqs. 8-11) to learn bulk motion with a low temporal resolution:

$$L_{II} = \mathbb{E}\left[||\mathcal{F}\left[\boldsymbol{I}_{ref}(\boldsymbol{x})\right] - \boldsymbol{s}(\boldsymbol{k})||_1\right] + \lambda_{TV} L_{TV} + \lambda_{MBC} L_{MBC} + \lambda_{ZMS} L_{ZMS} + \lambda_{Jac} L_{Jac}, \tag{15}$$

where $\lambda_{MBC}$, $\lambda_{ZMS}$, and $\lambda_{Jac}$ denote the weighting factors of the MBC loss, the zero-mean score loss, and the Jacobian loss, respectively. Specifically, MBCs $\boldsymbol{e}_i(\boldsymbol{x})$ (voxelized from MBC Gaussians) and MBC scores $\boldsymbol{w}_i(t)$ estimated from the MLP motion encoder are combined via Eq. 2 to obtain dynamic DVFs $\boldsymbol{d}(\boldsymbol{x}, t)$. By warping the reference-frame MRI $\boldsymbol{I}_{ref}$ using the DVFs, time-resolved dynamic MRIs are obtained and then converted to k-space to compare with the acquired k-space. In this stage, the in-plane image resolution was downsampled by a factor of two, while the through-plane resolution was preserved. In addition, for each stack, its two adjacent temporal neighbors were combined on-the-fly, resulting in a reduced effective temporal resolution. This coarse spatiotemporal training strategy emphasizes dominant bulk-motion patterns while suppressing high-frequency fluctuations and noise during early optimization. As a result, the model can first establish a stable low-resolution motion representation, which improves training robustness and reduces the risk of convergence to local minima.

In Stage III (**Fig. 2(c)**), the CNN motion encoder (Sec. 2.3.2) was incorporated and trained together with the reference-frame Gaussians (Eq. 6) and the MBC Gaussians (Eq. 4) to recover motion patterns at high spatiotemporal resolutions. At the beginning of this stage, Stage III(a) first initialized the CNN encoder by matching its motion scores to those predicted by the MLP motion encoder, which was trained in Stage II and frozen in Stages III and IV. The initialization objective was defined as:

$$L_{III}^{a} = MAE(Score_{MLP} - Score_{CNN}). \tag{16}$$

Following this warm start, Stage III(b) refined the full model using the native MRI spatial resolution and stack-wise temporal resolution with the objective:

$$L_{III}^{b} = \mathbb{E}\left[ ||\mathcal{F}\left[\boldsymbol{I}_{ref}(\boldsymbol{x})\right] - \boldsymbol{s}(\boldsymbol{k})||_1 \right] + \lambda_{TV} L_{TV} + \lambda_{MBC} L_{MBC} + \lambda_{ZMS} L_{ZMS} + \lambda_{Jac} L_{Jac} + \lambda_{Freq} L_{Freq}, \tag{17}$$

where $\lambda_{Freq}$ is the weighting factor for the frequency regularization term (Eq. 12). Compared with Stage II, this stage restores the original spatial and temporal resolutions of the acquisition, allowing the model to capture finer and more rapidly varying motion dynamics.

Finally, Stage IV served as a motion augmentation stage (**Fig. 2(d)**), in which the CNN-based motion encoder was fine-tuned using augmented motion patterns that were not observed during the training stages, thereby enhancing its robustness for real-time motion estimation. The objective was defined as the mean absolute error between the augmented MBC scores and the scores predicted by the CNN motion encoder:

$$L_{IV} = MAE\left(Score_{Aug} - Score_{CNN}\right). \tag{18}$$

Specifically, the solved MBC scores $\boldsymbol{w}_i(t)$ were randomly perturbed and applied to the corresponding MBCs to generate augmented DVFs, which were then used to deform the reference MRI volume and produce dynamic volumes with unseen motion patterns. The augmented volumes were subsequently processed with NUFFT to simulate raw k-space data, from which 2D FFT-based projection images were generated as inputs to the same motion encoder. The encoder was then trained to predict the perturbed MBC scores, and the predicted scores were compared with the target perturbed coefficients using a similarity loss. The perturbation of the MBC scores was defined as:

$$\boldsymbol{w'}_i(t) = r_1(t) \times r_2(i,k) \times \boldsymbol{w}_i(t), \tag{19}$$

where $r_1(t)$ acted as a global scaling factor shared across all motion levels $i$ and Cartesian directions $k$, thereby preserving inter-coefficient correlation within each motion instance, whereas $r_2(i,k)$ introduced additional level and direction-specific variations to increase motion diversity. During this stage, the reference image and MBC Gaussians were fixed, and only the CNN motion encoder was trained.

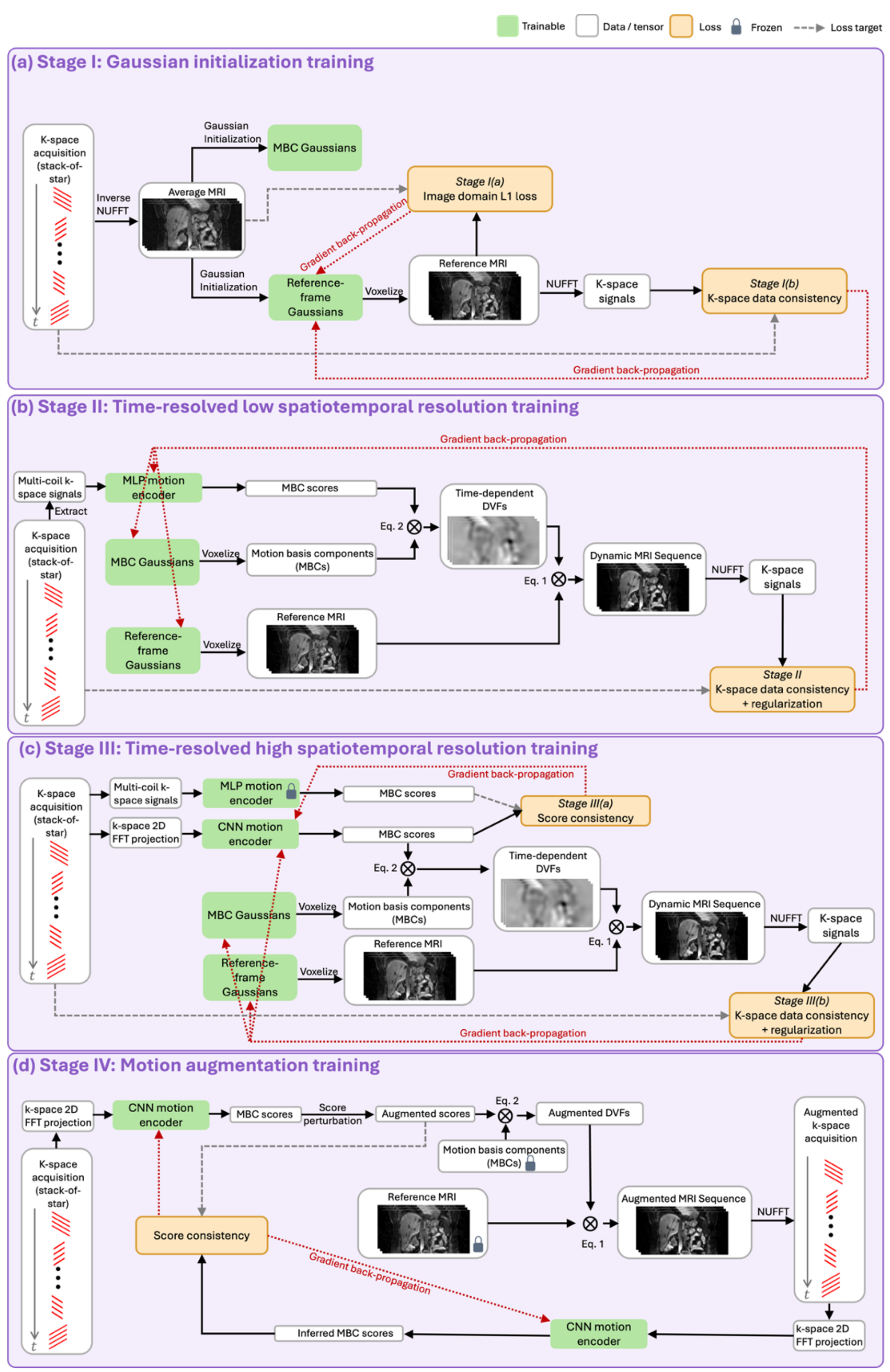

**Figure 2. Overview of staged training strategy.** (a) Stage I initializes the reference-frame MRI Gaussians from a motion-averaged reconstruction without modeling motion. (b) Stage II jointly optimizes the reference Gaussians, MBC Gaussians, and MLP encoder at reduced spatiotemporal resolutions to capture dominant bulk motion. (c) Stage III introduces the CNN encoder, first by matching the frozen MLP-derived scores and then by refining the full model at high spatiotemporal resolutions. (d) Stage IV freezes the learned reference image and MBCs and fine-tunes the CNN motion encoder using motion-augmented data to improve robustness to unseen motion states. Note that the two CNN motion encoders shown in this subfigure are identical and are displayed separately solely for visual clarity.

## 3. Experiments

*3.1 Evaluation datasets and metrics*

DREME-GSMR was evaluated using the Extended Cardiac Torso (XCAT) digital phantom (Segars *et al.*, 2010) simulations, physical phantom measurements, and clinical data. The XCAT simulation study provided 'ground-truth' anatomy and motion, enabling quantitative evaluation of both image reconstruction quality and motion estimation accuracy. The physical phantom experiments offered known motion trajectories under realistic acquisition conditions. Finally, the clinical dataset facilitated an assessment of DREME-GSMR's clinical potential and applicability.

**XCAT simulation study.** Six XCAT motion scenarios that differed in frequency, amplitude, and baseline patterns were generated. Scenario X1 modeled regular breathing with a 20 mm amplitude in superior-inferior (SI) direction and a 10 mm amplitude in anterior-posterior (AP) direction; X2 introduced a sudden baseline shift in the middle of the scan; X3 combined amplitude variations with a slow baseline drift; X4 featured a gradual reduction in breathing frequency with increasing amplitude; X5 simulated a slow breathing scenario; and X6 simultaneously varied baseline, amplitude, and frequency. A 30 mm-diameter liver tumor was inserted into the upper liver segment to serve as the tracking target. Because XCAT produces real-valued MR data, random phase angles were assigned to organs and tissues to simulate complex signals. Each motion scenario yielded 673 dynamic MR frames, equivalent to a 4.75-min scan with a 423 ms temporal resolution per SOS stack. To match realistic MR acquisition, the corresponding multi-coil k-space data were simulated after generating the ground-truth XCAT dynamic MRI volumes. A steady state spoiled gradient echo sequence with a TR of 4.5 ms was used. For each dynamic volume (423 ms temporal resolution), 94 spokes per stack were simulated in k-space following SOS golden-angle radial sampling, with each spoke containing 368 readout points and 8 receiver coils. Images were reconstructed with a volume size of $368 \times 368 \times 94$ at a $2 \times 2 \times 3\ mm^3$ voxel resolution.

The performance of the two learning tasks was separately evaluated. For dynamic MRI reconstruction (learning task 1), DREME-GSMR was trained separately on the six motion scenarios (X1-X6), and the image quality of the reconstructed dynamic MRI and the accuracy of the resolved respiratory motion were evaluated on the same motion scenarios. For real-time imaging (learning task 2), the model trained on a given motion scenario (e.g., X1) was cross-tested on the remaining scenarios (X2-X6) to assess its generalizability to unseen respiratory motion patterns. The image quality was evaluated using the structural similarity index measure (SSIM) (Zhou *et al.*, 2004). The motion estimation accuracy was evaluated by the center-of-mass error (COME) and the DICE similarity coefficient (DSC) of tracked tumor contours. The COME is defined as the Euclidean distance between the estimated and ground-truth centers-of-mass of the target. The DSC measures the spatial overlap between the estimated and ground-truth contours. Specifically, liver tumors were contoured from reference-frame MR images of each motion scenario and the tracking masks were then propagated to the dynamic/real-time MRI instances using the DVFs solved by DREME-GSMR. These propagated contours were then compared with the ground-truth tumor contours using COME and DSC.

**Physical phantom study**. A physical phantom study was conducted using a QUASAR MRI$^{4D}$ Motion Phantom, which consists of a water-filled outer body and a movable cylindrical insert and supports programmable motion along the SI direction. Similar to prior studies (Siddiq *et al.*, 2024), the imaging insert comprised an inner cylinder filled with a $CuSO_4$ solution and an air-filled spherical target mounted within the cylinder. Under T1-weighted imaging, the inner cylinder appeared bright while the spherical target and outer body were dark, providing clear contrast for motion evaluation. The inner cylinder was programmed to move along the SI direction in two irregular and two regular motion patterns. The two irregular motions (Motion 1 and Motion 2) had mean peak-to-valley displacements of 22.15 ± 2.32 mm over 95 cycles and 20.04 ± 4.59 mm over 47 cycles, respectively. The two regular motions (Motion 3 and Motion 4) were sinusoidal, with a period of 4 s and an amplitude of 30 mm, and a period of 3 s and an amplitude of 24 mm, respectively. Data were acquired on a 1.5 T MR-LINAC (Unity, Elekta AB, Stockholm, Sweden) at the UT Southwestern Medical Center with a TR of 4.5 ms and 8 receive coils. A total of 673 stacks were continuously acquired using SOS golden-angle radial sampling for ~5 minutes. The reconstructed images had a spatial resolution of $2 \times 2 \times 3\ mm^3$ with a matrix size of $368 \times 368 \times 94$.

Similar to the XCAT study, for dynamic MRI reconstruction, DREME-GSMR was trained separately for each of the four motion scenarios, and the accuracy of the solved motion was evaluated within the corresponding scenario. For real-time imaging, a cross-validation scheme was used to assess motion estimation accuracy. Specifically, the spherical target inside the phantom was contoured, and its motion was compared with the programmed ground-truth curves by COME.

**Clinical study.** Clinical data were acquired on the same 1.5 T MR-LINAC as the physical phantom study, from 6 healthy volunteers and 20 patients. The volunteers were imaged under free-breathing conditions, whereas the patient cases were acquired under their motion-management protocols (free-breathing with or without abdominal compression). T1-weighted images were obtained with a matrix size of $500 \times 500 \times (61 - 79)$ and a spatial resolution of $1.5 \times 1.5 \times 3\ mm^3$, acquired with a TR of 5.1 ms and 8 receive coils. A total of 878-921 radial stacks were continuously acquired for each scan, corresponding to a temporal resolution of ~400 ms per stack and a scan time of ~6 mins.

For clinical data, where no absolute ground truth exists, each subject had two sequential scans (A/B) to enable cross-scan validation: the model trained on scan A was used to infer real-time motion/imaging on scan B, and evaluated against the dynamic reconstruction results of a reference model trained on scan B (and vice versa). For each case, the liver was contoured, and its COME was calculated between the real-time prediction and the corresponding 'pseudo-ground-truth' dynamic reconstruction. This evaluation strategy was designed to assess the generalizability of the proposed framework in a realistic clinical setting, where a pre-treatment scan is available for model training, whereas real-time inference must be performed on online MR signals acquired under potentially different motion patterns.

## 3.2 Implementation details

Implementation details of DREME-GSMR are as follows: (a) DREME-GSMR was implemented in PyTorch 1.13.1 (Paszke *et al.*, 2019), and trained with the Adam optimizer (Kingma and Ba, 2014); (b) The NUFFT operator was adopted from the TorchKbNufft library (Muckley *et al.*); (c) For the XCAT and clinical studies, readouts at the three $k_x = k_y = 0$ central locations around $k_z = 0$ were used as input for the MLP-based motion encoder. In the physical phantom study, nine central $k_z$ locations were used to better capture the locally restricted motion of the phantom; (d) Due to the wide variations of raw k-space signals $\boldsymbol{s}(\boldsymbol{k}, t)$ across different coils, z-score normalization was applied to the real and imaginary channels before inputting them into the motion encoder; (e) Regarding DREME-GSMR's Gaussian initializations, the number of cloud points is set as $M = 100{,}000$ to initialize reference-frame Gaussians, and for the three levels of MBC Gaussians, $M_1 = 15^3, M_2 = 20^3$, and $M_3 = 25^3$ were set, respectively, to match with the number of B-spline control points in prior studies (Shao *et al.*, 2025b; Shao *et al.*, 2024b); (f) Weighting factors were empirically set to $\lambda_{TV} = 2 \times 10^{-6}$, $\lambda_{MBC} = 1 \times 10^{-2}$, $\lambda_{ZMS} = 1 \times 10^{-4}$, $\lambda_{Jac} = 1 \times 10^{-1}$, and $\lambda_{Freq} = 1$; (g) The numbers of epochs for Stage I(a), Stage I(b), Stage II, Stage III(a), Stage III(b), and the motion augmentation stage IV were 400, 300, 1200, 5000, 1500, and 1000, respectively; (h) The learning rate of the MLP motion encoders was set to $2 \times 10^{-3}$ and $5 \times 10^{-4}$ for the physical phantom data, and the clinical/XCAT datasets, respectively. The learning rate of the CNN was set to $2 \times 10^{-3}$ at Stage III(a) and changed to $1 \times 10^{-4}$ for later stages; and (i) The motion augmentation factors were set to $r_1 \in [0.2, 2.6]$ and $r_2 \in [0.4, 1.6]$. For the physical phantom study, in which motion was restricted to the SI direction, motion augmentation was applied only along that direction.

## 3.3 Comparison and ablation studies

### 3.3.1 Comparison methods

DREME-GSMR was compared with XD-GRASP (Feng *et al.*, 2016), a PCA-based motion modeling method, and the DREME-MR model (Shao *et al.*, 2025b), which is considered the state-of-the-art. XD-GRASP is an established respiratory motion-resolved reconstruction framework that reconstructs amplitude-sorted motion states rather than dynamic volumes. Implemented with 10 respiratory bins, it cannot fully represent irregular or nonperiodic motion and does not support real-time imaging. Therefore, XD-GRASP was used to highlight the advantage of dynamic MRI in capturing motion variations that cannot be fully described by amplitude-binned 4D-MRI.

Because no publicly available framework jointly performs patient-specific dynamic reconstruction and real-time volumetric inference in a single pipeline, a PCA-based baseline was built on top of XD-GRASP to support both tasks for comparison with DREME-GSMR. Specifically, a pre-treatment MRI acquisition was amplitude-sorted into 10 respiratory bins and reconstructed into a 4D-MRI using XD-GRASP, and PCA was applied to its inter-bin DVFs (relative to the end-exhale reference bin) to calculate patient-specific motion bases. With these frozen PCA bases and the same pre-treatment MRI raw data, an MLP motion encoder was trained to estimate PCA coefficients from k-space signals while re-fining the reference-bin image. Conceptually, this baseline is similar to STINR-MR (Shao *et al.*, 2024a) with modifications that enable real-time inference (via the MLP motion encoder). It differs from DREME-GSMR in that its motion bases are predefined by PCA rather than learned jointly with the reconstruction, and it uses an MLP-only encoder without motion augmentation. For comparison with DREME-MR (Shao *et al.*, 2025b), the original architecture and settings of DREME-MR were kept, but the cardiac-motion components were removed and the multi-resolution training was enabled (Sec. 2.3.4) for a fair comparison.

### 3.3.2 Ablation and robustness studies

Two ablation studies were conducted to analyze key design choices in DREME-GSMR. To evaluate the benefit of the dual-path motion encoder, DREME-GSMR was compared with two single-encoder variants, denoted DREME-GSMR-MLP and DREME-GSMR-CNN, in which only the MLP-based or CNN-based motion encoder was used under the same overall training framework. Notably, for DREME-GSMR-MLP, the same motion augmentation strategy (**Fig. 2**) was used to ensure a fair comparison, except that simulated central k-space signals rather than 2D FFT-based projection images were used as encoder inputs.

The robustness of DREME-GSMR was also evaluated to variations in scan time of the pre-treatment training MRI set. Because the number of radial stacks in the training set determines both the signal-to-noise ratio (SNR) and streak-artifact level of the reconstructed image, retrospectively reducing the scan time (and thus the sampled number of stacks) degrades image quality without altering the underlying anatomy. Apparent SNR was computed from magnitude reference-frame reconstructions as $\mathrm{SNR} = 0.655 \cdot \overline{S_{\mathrm{ROI}}}/\sigma_{\mathrm{bg}}$, where $\overline{S_{\mathrm{ROI}}}$ is the mean intensity within the contoured ROI (target region in the phantom and liver in the patient), $\sigma_{\mathrm{bg}}$ is the standard deviation within a fixed signal-free background ROI, and the factor 0.655 corrects for the Rayleigh distribution of magnitude background noise (Yu *et al.*, 2018; Dietrich *et al.*, 2007; Gudbjartsson and Patz, 1995; Kaufman *et al.*, 1989). Because this background-noise-based SNR does not capture coherent undersampling streak artifacts, we additionally report SSIM and peak SNR (PSNR) of each reconstruction relative to the full-data reconstruction. Specifically, the DREME-GSMR model was trained using 75%, 50%, 25%, and 20% of the original radial stacks and the results were compared with the full-data model (100%). Three evaluation settings were considered: (1) **Same-scenario/scan dynamic reconstruction**: the model was trained and tested on the same scenario/scan using the specified sampling ratio. (2) **Same-scenario/scan real-time imaging**: the model was trained on the sampled portion of a scenario/scan and then used to estimate motion in the remaining frames of the same scenario/scan. Consequently, the 100% sampling condition is not applicable because no frames remain for testing. (3) **Cross-scenario/scan real-time imaging**: the model was trained on one scenario/scan using the specified sampling ratio and tested on a fully sampled motion scenario or paired scan.

## 4. Results

### *4.1 The XCAT study results*

In the XCAT study, DREME-GSMR improved reference-frame MRI reconstruction quality compared with DREME-MR across all six motion scenarios. As shown in **Fig. 3** and **Table I**, both models reduced artifacts and motion-induced blurring substantially, but DREME-GSMR produced sharper anatomical boundaries and achieved a higher SSIM than DREME-MR for same-scenario dynamic reconstruction (0.92 ± 0.01 vs. 0.84 ± 0.02). The zoomed-in panels show sharper structural boundaries, suggesting that the explicit Gaussian representation preserves high-frequency details that the INR-based DREME-MR tends to over-smooth.

DREME-GSMR also maintained high motion-tracking accuracy in both dynamic reconstruction and cross-scenario real-time imaging (**Fig. 4**). When trained on the X1 motion scenario with regular breathing cycle/amplitude and tested on all scenarios (X1-X6), both DREME-GSMR and DREME-MR faithfully captured the motion trends. Quantitatively, DREME-GSMR achieved consistently lower COME (**Table I**) than DREME-MR while using 70% of the training time (~50 vs. ~70 min on an NVIDIA 4090 GPU). Replacing the INR/B-spline representation with a 3D Gaussian representation thus improved image quality and motion accuracy simultaneously.

**Figure 5** shows representative dynamic MRI reconstructions by DREME-GSMR for motion scenario X2 at four time points. The reconstructed SI and AP motion trajectories closely match the ground truth, and the axial, coronal, and sagittal views demonstrate good agreement between the reconstructed and ground-truth XCAT volumes. In particular, the deformation of the liver and tumor is well preserved, with minor errors primarily around anatomical boundaries.

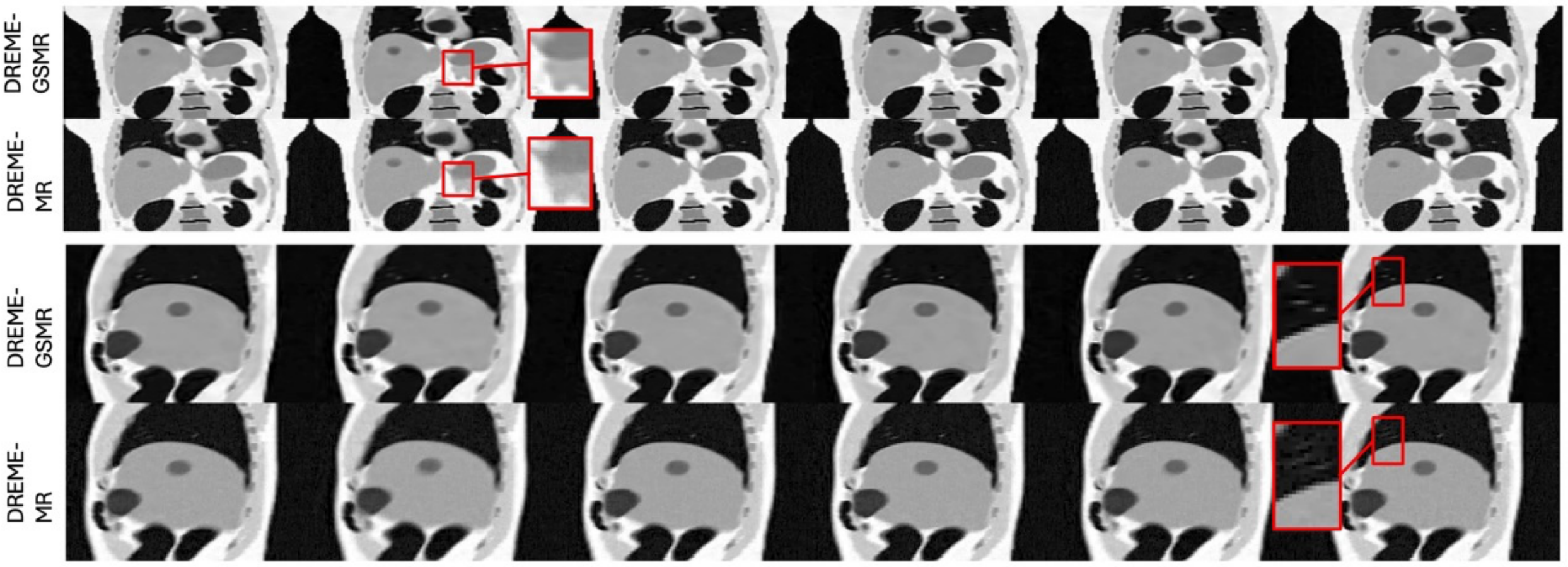

**Figure 3.** Comparison of reconstructed reference-frame MRIs from six motion scenarios (X1-X6, from left to right) between DREME-MR and DREME-GSMR.

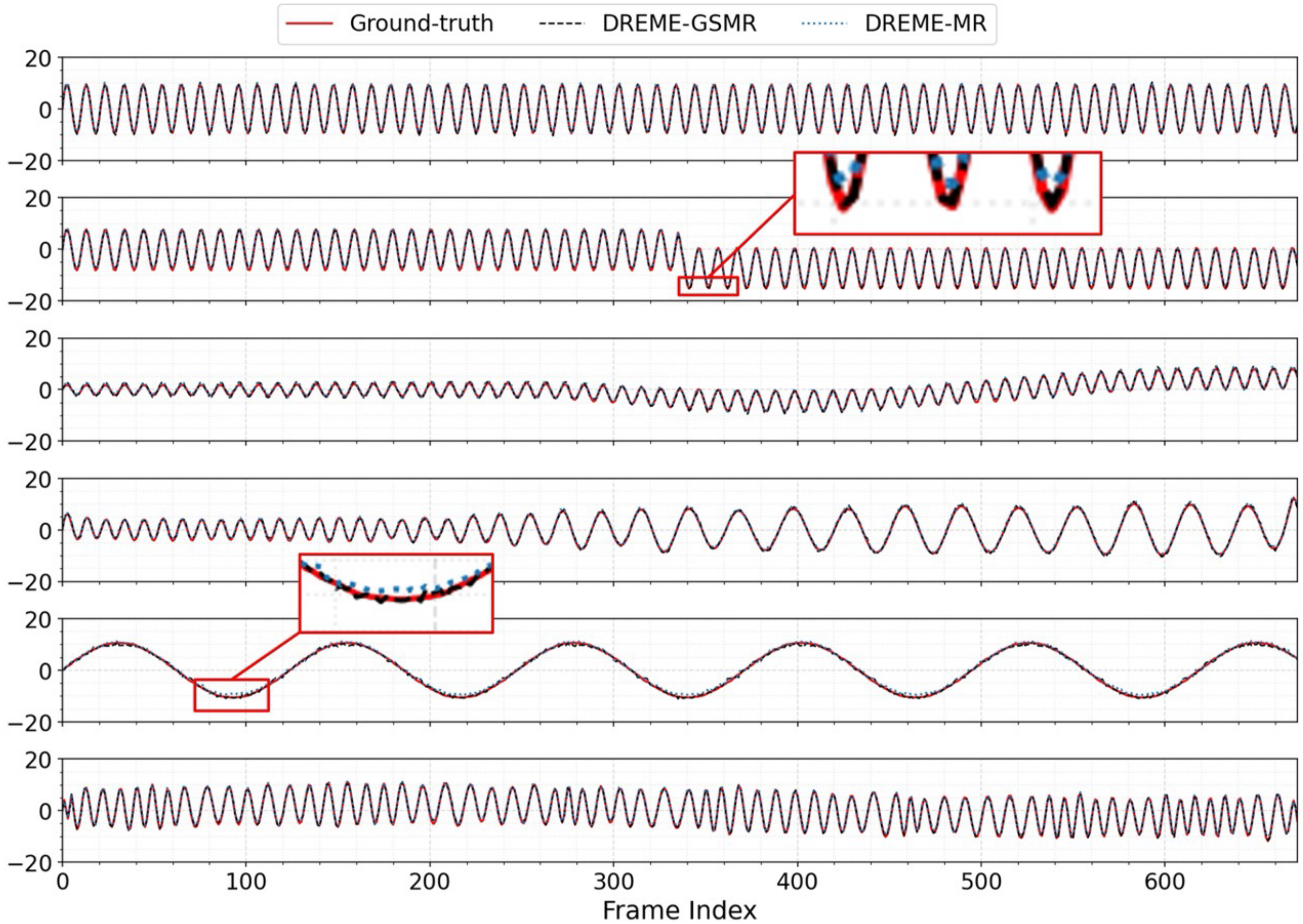

**Figure 4.** Liver tumor center-of-mass trajectories in the SI direction of the XCAT study. The DREME-MR and DREME-GSMR were trained on the X1 scenario and tested on all six (X1-X6) scenarios.

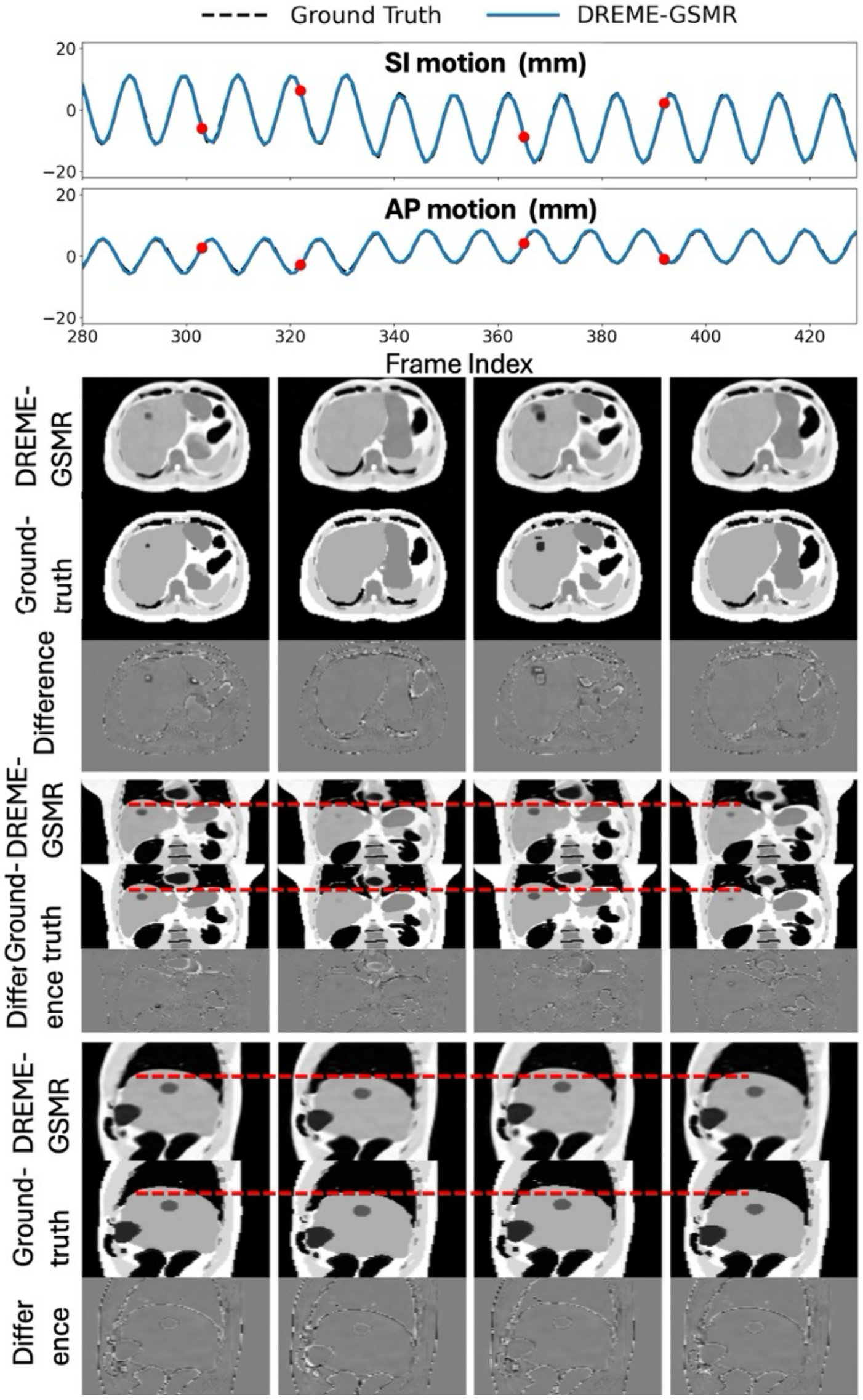

**Figure 5.** An example of DREME-GSMR's dynamic reconstruction results in the XCAT study (motion scenario X2). The first two rows show the estimated motion curves along the SI and AP directions. The following rows compare the reconstructed MR images with the ground-truth images at four time points at three views. The window widths of the difference images are half of those of the MR images.

**Table I.** Accuracy of the solved dynamic MRIs and motion for the XCAT study. The tasks of dynamic reconstruction and real-time imaging are separately evaluated. The results are presented as the mean ± standard deviation. Better values are in bold. The arrows are pointing in the direction of higher accuracy.

| Method | Same-scenario dynamic reconstruction | | |
|---|---|---|---|
| | DSC↑ | SSIM↑ | COME (mm)↓ |
| DREME-MR | **0.92±0.02** | 0.84±0.02 | 0.66±0.21 |
| DREME-GSMR | **0.92±0.02** | **0.92±0.01** | **0.50±0.15** |
| Method | Cross-scenario real-time imaging | | |
| | DSC↑ | SSIM↑ | COME (mm)↓ |
| DREME-MR | 0.91±0.02 | 0.82±0.02 | 0.74±0.27 |
| DREME-GSMR | **0.92±0.03** | **0.91±0.02** | **0.65±0.19** |

*4.2 The physical phantom study results*

In the physical phantom study, DREME-GSMR achieved the most accurate dynamic motion reconstruction across all four programmed motion scenarios. As summarized in **Table II**, the mean target COME was 1.20 mm, compared with 3.04 mm for XD-GRASP and 2.50 mm for the PCA-based motion modeling method. The trajectory comparisons in **Fig. 6(a)** show that DREME-GSMR better preserved motion amplitude and cycle-to-cycle variability, particularly under irregular motion. XD-GRASP exhibited the largest deviations, as its amplitude-sorting strategy represents motion using only a limited number of motion states and is therefore less capable of capturing irregular temporal variations, whereas PCA showed intermediate performance. For real-time imaging, models were trained on Motion 4 and applied to unseen Motion 1 (**Fig. 6(b)**). XD-GRASP was not included because it was not designed for real-time motion estimation. DREME-GSMR followed the ground-truth trajectories more consistently than PCA, which shows clear motion undershooting/overshooting in the zoomed-in panels, and achieved lower COME for all cross-scenario real-time imaging experiments (**Table II**). **Figure 7** shows an example of DREME-GSMR-resolved dynamic MRI sequence for Motion 1. The reconstructed SI trajectory closely matches the programmed ground-truth motion, and the sagittal images at the four selected time points clearly demonstrate the displacement of the inner cylinder relative to the fixed reference line.

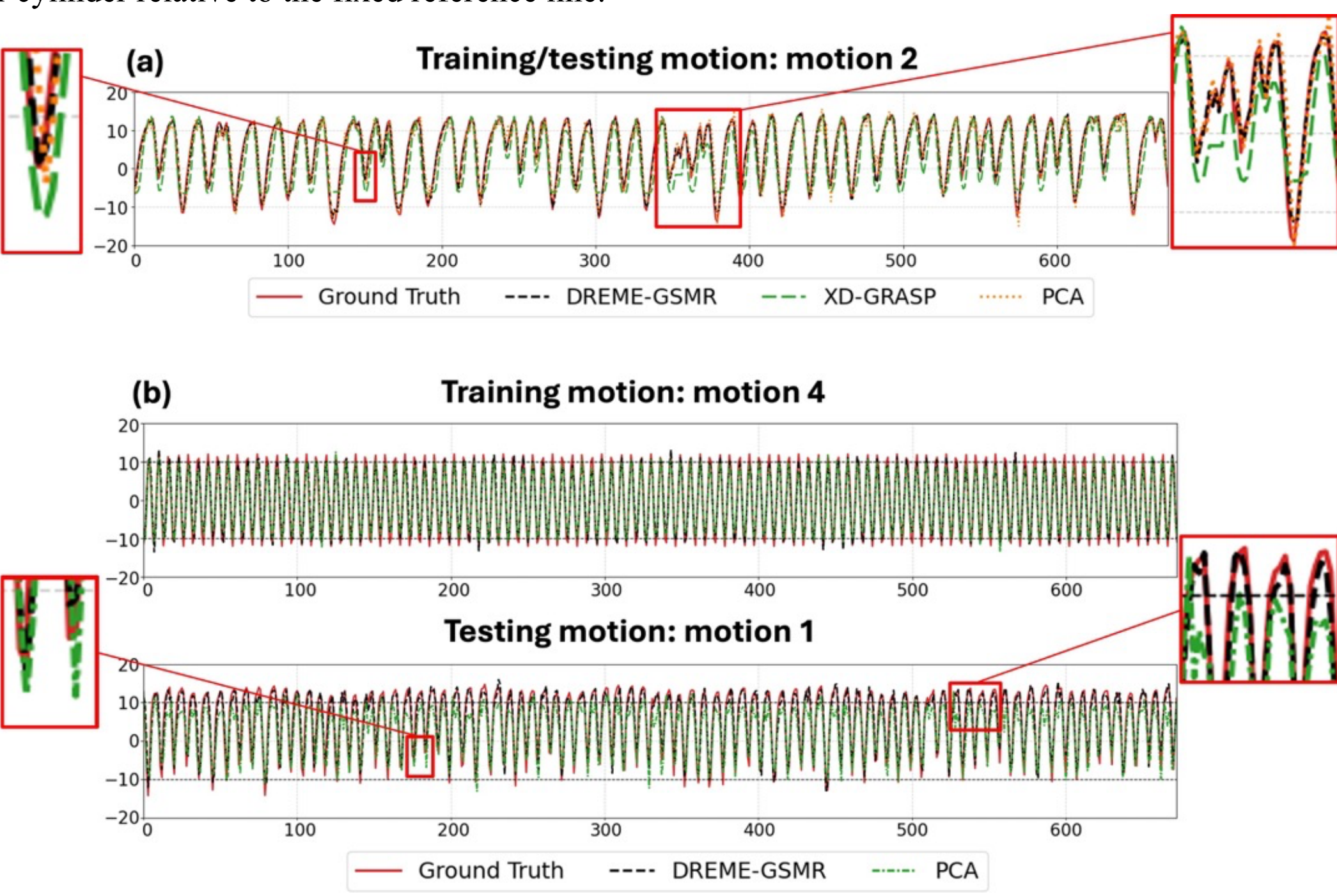

**Figure 6.** (a) SI center-of-mass trajectories of the spherical target in the physical phantom study for the dynamic reconstruction of Motion 2. Trajectories from DREME-GSMR, XD-GRASP, and PCA reconstructions are overlaid with the programmed ground-truth curves. Enlarged views highlight representative regions with visible differences among the methods. (b) Real-time motion estimation trajectories in the physical phantom study. DREME-GSMR and PCA were trained on Motion 4. The top row shows the training motion, and the lower three rows show testing on unseen Motion 1. The horizontal dashed lines indicate SI positions of -10 and +10 mm.

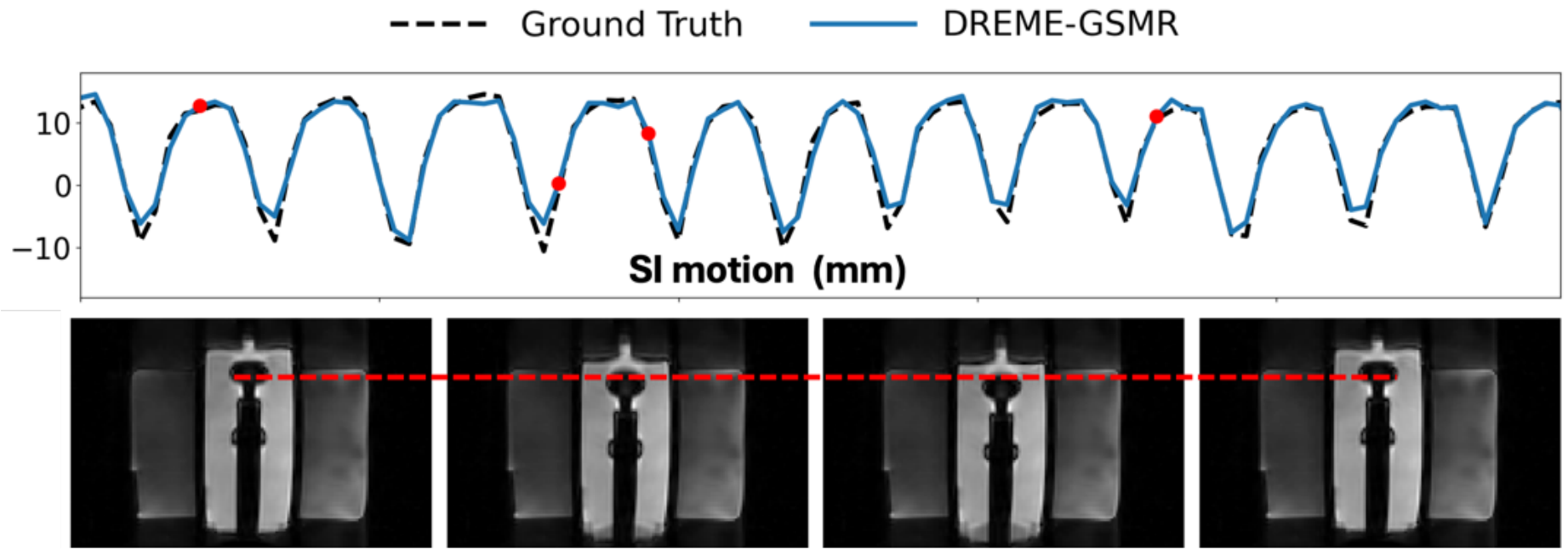

**Figure 7.** Dynamic MR images of one phantom case (Motion 1) reconstructed by DREME-GSMR. The top row compares the reconstructed SI motion trajectory with the programmed ground-truth curve, and the red markers indicate the four time points displayed below. The second row shows sagittal images at these four time points.

**Table II.** COME of the spherical target for same-scenario dynamic reconstruction and cross-scenario real-time imaging in the physical phantom study. Results are reported as mean ± standard deviation (mm). Better values are in bold.

| Motion Scenarios | Same-scenario dynamic reconstruction | | |
|---|---|---|---|
| | DREME-GSMR | XD-GRASP | PCA |
| Motion 1→1 | **1.03±1.04** | 4.71±3.61 | 2.90±2.07 |
| Motion 2→2 | **0.87±0.71** | 3.29±2.18 | 2.07±1.54 |
| Motion 3→3 | **1.52±0.86** | 2.08±1.35 | 3.15±1.83 |
| Motion 4→4 | **1.36±0.96** | 2.08±1.63 | 1.88±1.06 |
| Motion Scenarios | Cross-scenario real-time imaging | | |
| | DREME-GSMR | XD-GRASP | PCA |
| Motion 1→2,3,4 | **1.41±1.16** | N/A | 3.45±2.22 |
| Motion 2→1,3,4 | **1.73±1.38** | N/A | 3.84±2.36 |
| Motion 3→1,2,4 | **1.27±1.02** | N/A | 3.43±2.30 |
| Motion 4→1,2,3 | **1.41±1.16** | N/A | 2.92±1.89 |

*4.3 The clinical study results*

In the clinical study, DREME-GSMR maintained low real-time liver localization error under sequential-scan cross-evaluation. Because the volunteers were imaged under free-breathing conditions, whereas the patients were scanned under motion-management protocols (free-breathing with or without abdominal compression), the two groups were analyzed separately (full results in **Appendix: Table A.1** and **Appendix: Fig. A.1**). Volunteers exhibited substantially larger respiratory motion than the patients, with an average SI peak-to-valley motion range of 10.26 ± 5.61 mm versus 4.47 ± 2.63 mm (**Table III**). Using the dynamic reconstruction results as 'pseudo ground truth', DREME-GSMR achieved a mean 3D liver localization error of 1.31 ± 0.82 mm for the volunteer group and 0.96 ± 0.64 mm for the patient group, maintaining accurate real-time motion estimation in both free-breathing and motion-managed clinical settings. Representative cross-scan real-time liver SI trajectories predicted by DREME-GSMR from three volunteers and three patients closely matched the 'pseudo ground truth' (**Fig. 8**), and a representative clinical example showed good visual agreement between the predicted and 'pseudo-ground-truth' MRI images across all three views (**Fig. 9**).

A comparison with XD-GRASP and PCA is summarized in **Table IV**. Because no ground truth was available in the clinical study, XD-GRASP and PCA dynamic reconstructions were compared against the DREME-GSMR dynamic reconstruction from the same scan, which served as the reference. PCA agreed more closely with DREME-GSMR than XD-GRASP, which more frequently underestimated irregular motions, as illustrated for case P3 in **Fig. 10(a)**, consistent with observations made in the physical phantom study. For real-time imaging, DREME-GSMR consistently achieved lower localization errors than PCA (**Table IV**). In case V4, where scan A and scan B had large differences in SI motion ranges (10.87 ± 1.05 mm and 4.06 ± 0.78 mm, respectively), PCA failed to generalize to the unseen motion amplitude (arrow-highlighted region) while DREME-GSMR remained closely aligned with the 'pseudo-ground-truth' trajectory (**Fig. 10(b)**), demonstrating better robustness to cross-scan motion variations.

**Table III.** Average peak-to-valley respiratory motion range and cross-scan real-time liver localization error (by DREME-GSMR) in the clinical dataset. Results are reported separately for healthy volunteers and patients and are presented as COME in mean ± standard deviation (mm).

| Dataset | Motion Range (mm) | | | Localization Error (mm) | | | |
|---|---|---|---|---|---|---|---|
| | SI | AP | LR | 3D | SI | AP | LR |
| Volunteer | 10.26±5.61 | 3.33±2.03 | 0.91±0.58 | 1.31±0.82 | 0.83±0.74 | 0.57±0.57 | 0.47±0.40 |
| Patient | 4.47±2.63 | 1.98±1.43 | 0.56±0.36 | 0.96±0.64 | 0.60±0.53 | 0.49±0.50 | 0.33±0.31 |

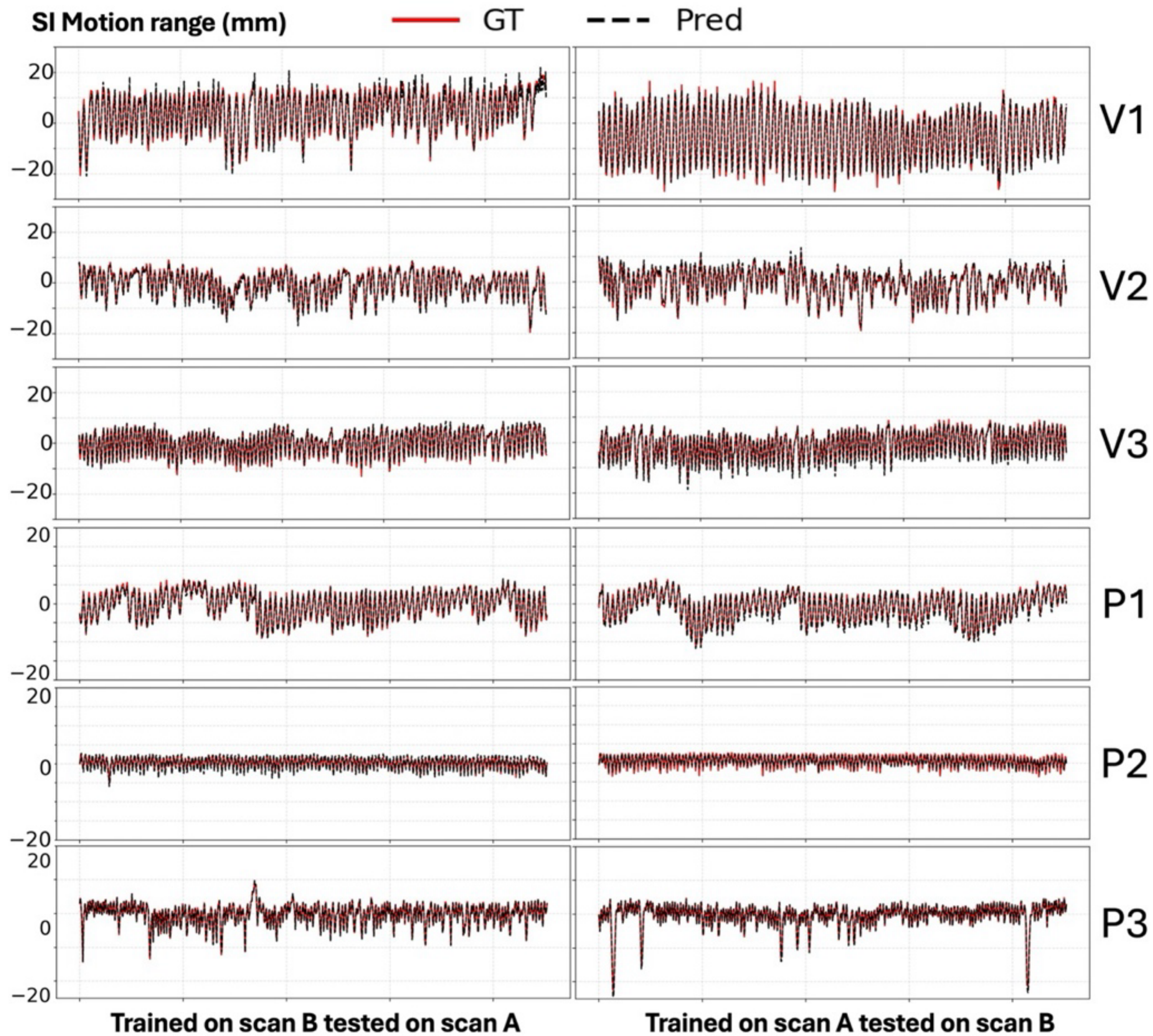

**Figure 8.** Representative liver SI motion trajectories from clinical scans of volunteers (V1-V3) and patients (P1-P3) under cross-scan testing. For each case, the DREME-GSMR model was trained on one scan and tested on the paired scan, and the dynamic reconstruction of the test scan was used as the 'pseudo ground truth'. The two columns show the two directional cross-scan testing experiments between the paired A/B scans. Left: trained on scan B tested on scan A. Right: trained on scan A tested on scan B.

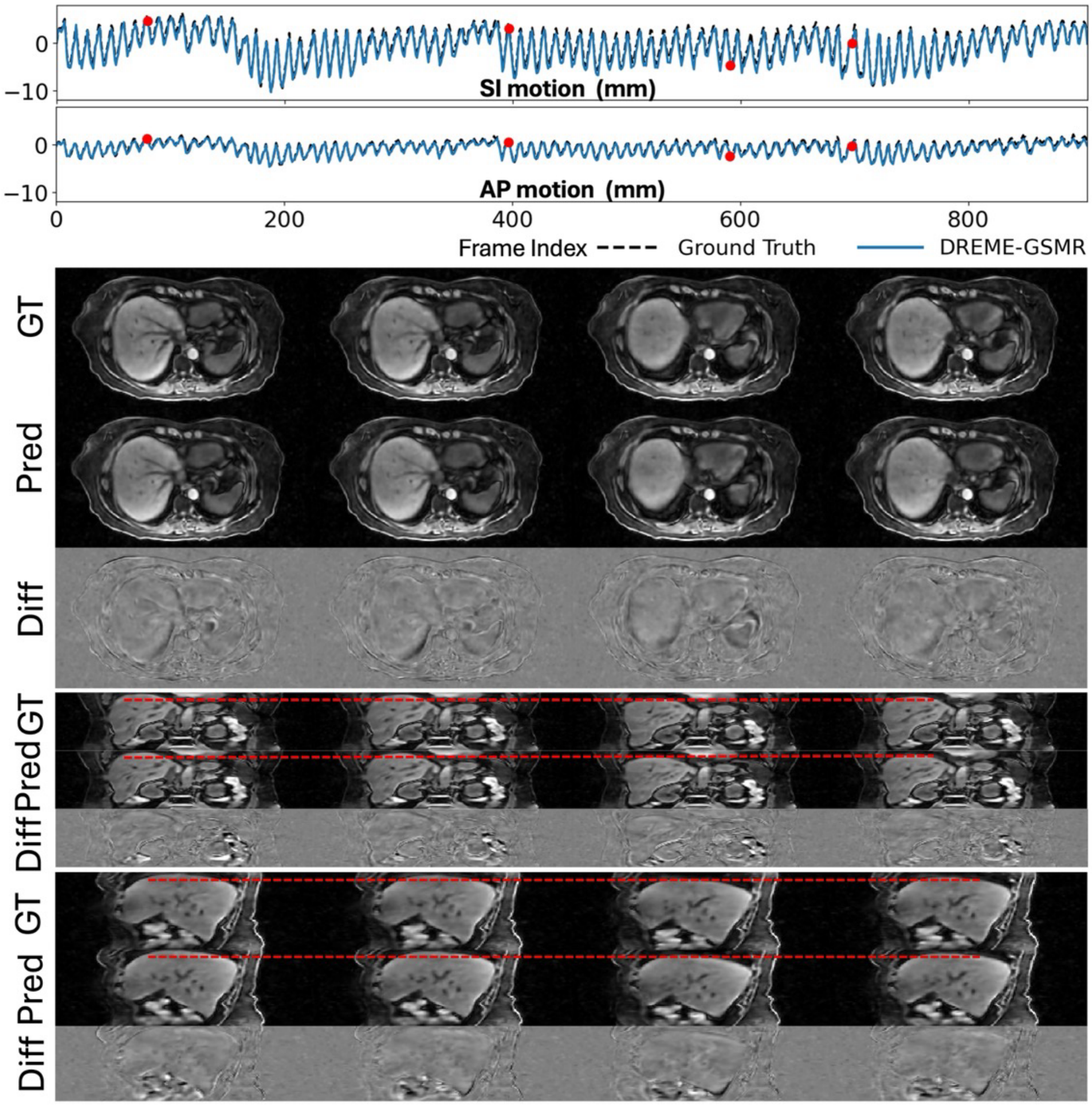

**Figure 9.** Example of reconstructed clinical scans. DREME-GSMR was trained on scan A of P1, and tested on scan B for real-time prediction, using scan B's dynamic reconstruction as 'pseudo ground truth'. The top two rows show the SI and AP motion trajectories, with red markers indicating the four time points displayed below. The remaining rows show axial, coronal, and sagittal views of the 'pseudo ground truth', DREME-GSMR prediction, and difference images at the selected time points.

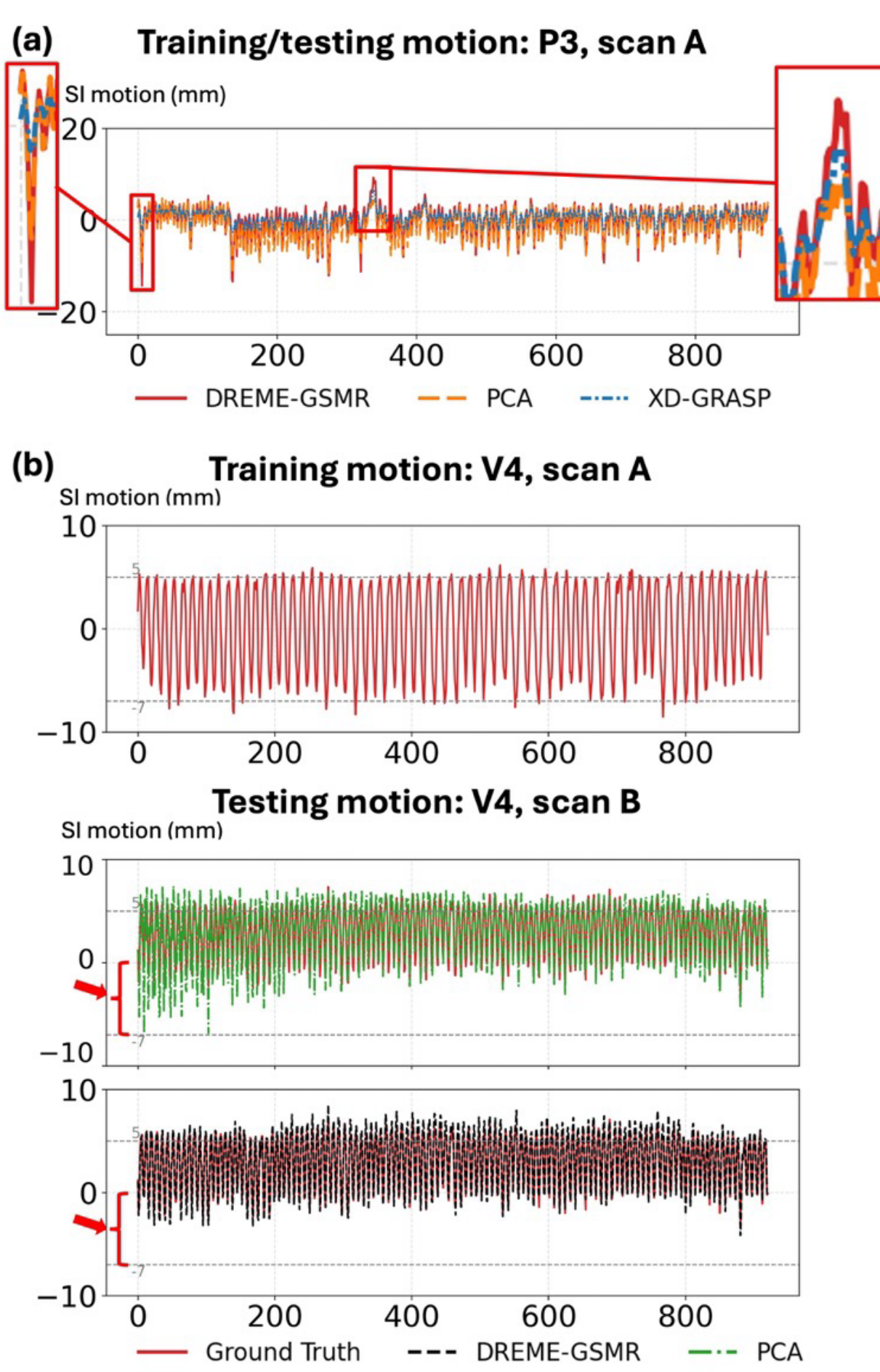

**Figure 10.** (a) Comparison of liver center-of-mass SI trajectories derived from dynamic reconstructions for case P3. DREME-GSMR, PCA, and XD-GRASP reconstructions from the same scan are overlaid to illustrate the differences in solved motion patterns. (b) Comparison of cross-scan real-time SI motion estimation for V4. Both DREME-GSMR and PCA were trained on scan A and tested on scan B, and the dynamic reconstruction of scan B was used as 'pseudo ground truth'. The top row shows the training motion from scan A, and the middle and bottom rows show the PCA and DREME-GSMR predictions on scan B, respectively. Horizontal dashed lines indicate reference SI positions.

**Table IV.** Comparison of DREME-GSMR, XD-GRASP, and PCA for clinical dynamic reconstruction and cross-scan real-time imaging. Results are reported for the clinical subset cases V1, V4, P1, and P3 as COME in mean ± standard deviation (mm). Lower values indicate better performance.

| Evaluation setting | DREME-GSMR | XD-GRASP | PCA |
|---|---|---|---|
| Same-scan dynamic reconstruction | N/A | 1.95±1.45 | 1.58±1.00 |
| Cross-scan real-time imaging | 1.11±0.98 | N/A | 1.98±1.48 |

*4.4 The ablation study results*

*4.4.1 Encoder design comparison*

The dual-path encoder outperformed both single-encoder variants (DREME-GSMR-MLP and DREME-GSMR-CNN), achieving the lowest errors across all phantom and clinical experiments for both dynamic reconstruction and cross-scenario/scan real-time imaging (**Table V**). Although all three models followed the training motion reasonably well, the two single-encoder variants showed reduced robustness on unseen motions: in the physical phantom study, both variants less accurately recovered peak motion amplitudes in the enlarged views, and in a representative clinical case (P1 scan B), both exhibited motion undershooting and mismatch relative to the 'pseudo ground truth', whereas DREME-GSMR remained more closely aligned (**Appendix: Fig. A.2**).

**Table V.** Comparison of DREME-GSMR and the two single-encoder variants in the physical phantom and clinical study. Results are reported for four phantom cases and the clinical subset cases V1, V4, P1, and P3 as COME in mean ± standard deviation (mm) for dynamic reconstruction and cross-scenario/scan real-time imaging. Lower values indicate better performance.

| Dataset | Evaluation setting | DREME-GSMR | DREME-GSMR-MLP | DREME-GSMR-CNN |
|---|---|---|---|---|
| Physical phantom | Same-scenario dynamic reconstruction | **1.20±0.94** | 2.07±1.56 | 1.34±1.10 |
| Physical phantom | Cross-scenario real-time imaging | **1.45±1.20** | 2.40±1.59 | 1.82±1.75 |
| Clinical | Cross-scan real-time imaging | **1.11±0.98** | 1.81±1.36 | 2.04±2.06 |

*4.4.2 Undersampling study results*

The performance of DREME-GSMR degraded as the sampling ratio decreased (**Tables VI and VII**). In the phantom study, the mean dynamic-reconstruction error rose only from 1.20 mm at 100% to 1.45 mm at 25% before increasing more sharply at 20% (1.84 mm), with the same pattern for same- and cross-scenario real-time imaging (**Appendix: Fig. A.3(a)**). The clinical study followed the same trend (**Table VII**), where the model performance was stable from 75% to 25% but degraded markedly at 20%, particularly for cross-scan testing (**Appendix: Fig. A.3(b)**). Reference-frame image quality trended similarly (**Fig. 11, Table VI**): SSIM, PSNR, and SNR decreased monotonically with sampling ratio, with 50-75% reconstructions visually comparable to the full-data result and clear artifacts appearing only at 20%. Although the SNR of NUFFT reconstructions fell steeply (9.72 at 100% to 5.42 at 20% for case P1), the DREME-GSMR reference-frame MRI retained substantially higher SNR at every ratio (21.14 and 8.24, respectively).

Overall, DREME-GSMR was robust to the noise and aliasing of aggressive undersampling, maintaining dynamic-reconstruction and real-time motion errors within ~2 mm using 25-50% of the data, with a noticeable breakdown at 20%. For a typical clinical scan time of ~6 min, this corresponds to a potential reduction in acquisition time to about 1.5-3 min, which may facilitate a more efficient treatment workflow.

**Table VI.** SSIM, PSNR, and SNR of physical phantom and clinical reference-frame MRIs reconstructed under different undersampling ratios, using the full-acquisition reconstruction as reference (for SSIM and PSNR). Results are reported for four phantom cases and the clinical subset cases V1, V4, P1, and P3 as mean ± standard deviation.

| Undersample ratio | Phantom | | | Clinical | | |
|---|---|---|---|---|---|---|
| | SSIM | PSNR | SNR | SSIM | PSNR | SNR |
| 20% | 0.73±0.01 | 34.20±0.29 | 7.04±0.61 | 0.61±0.08 | 29.06±2.52 | 8.36±1.55 |
| 25% | 0.84±0.06 | 36.00±0.32 | 8.79±0.73 | 0.75±0.06 | 31.05±2.13 | 13.16±1.41 |
| 50% | 0.93±0.05 | 37.60±0.33 | 14.12±0.93 | 0.86±0.03 | 33.30±1.71 | 17.86±1.58 |
| 75% | 0.94±0.03 | 37.61±0.19 | 17.16±1.40 | 0.89±0.02 | 34.92±1.61 | 23.18±2.35 |
| 100% | NA | NA | 19.11±2.16 | NA | NA | 23.39±1.70 |

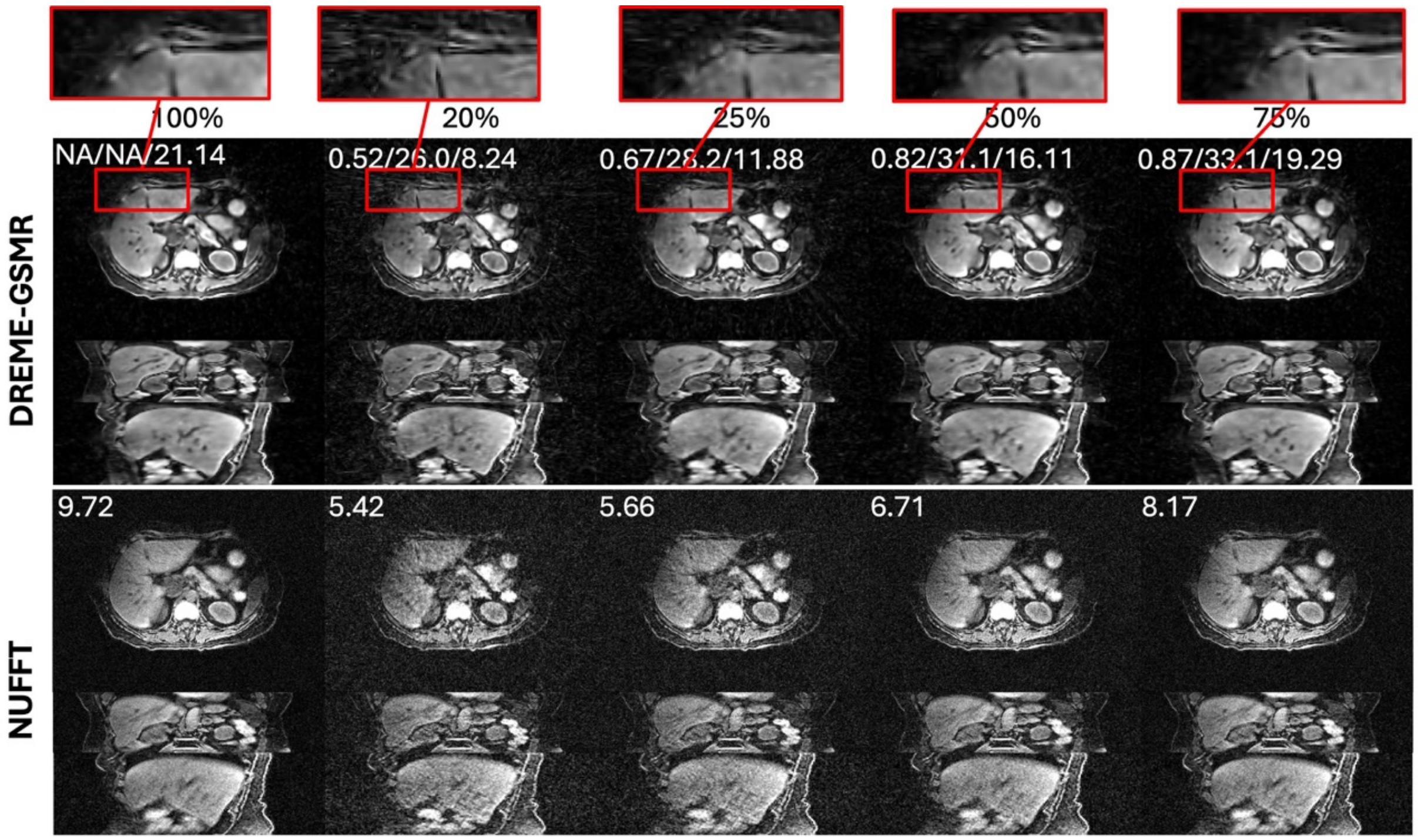

**Figure 11.** Reference-frame images reconstructed by DREME-GSMR and NUFFT from the clinical undersampling study (P1, scan A) at 100%, 20%, 25%, 50%, and 75% sampling ratios. For DREME-GSMR reconstructions, numbers displayed on the undersampled panels denote SSIM/PSNR (relative to the 100% reconstruction) and SNR. For NUFFT reconstructions, numbers displayed denote SNR. Enlarged views highlight representative image degradation caused by undersampling.

**Table VII.** Physical phantom and clinical undersampling study results. Errors are reported for four phantom cases and the clinical subset cases V1, V4, P1, and P3 as COME in mean ± standard deviation (mm) for same-scenario/scan dynamic reconstruction, same-scenario/scan real-time imaging, and cross-scenario/scan real-time imaging under different sampling ratios. For the same-scenario/scan dynamic reconstruction results, the physical phantom study used the programmed ground-truth motion curves as reference, while the clinical study used the reconstructed motion curves by 100% sampled data as reference. Thus, only the physical phantom can be evaluated for 100% sampling. Lower values indicate better performance.

| Dataset | Same-scenario/scan dynamic reconstruction | | | | |
|---|---|---|---|---|---|
| | 100% | 75% | 50% | 25% | 20% |
| Physical phantom | 1.20±0.94 | 1.27±1.02 | 1.41±1.17 | 1.45±1.22 | 1.84±1.45 |
| Clinical | N/A | 1.04±0.60 | 1.25±0.77 | 1.48±0.77 | 2.34±1.51 |
| Dataset | Same-scenario/scan real-time imaging | | | | |
| | 100% | 75% | 50% | 25% | 20% |
| Physical phantom | N/A | 1.39±1.11 | 1.46±1.23 | 1.62±1.29 | 2.14±1.49 |
| Clinical | N/A | 1.14±0.66 | 1.43±0.88 | 1.68±0.84 | 2.63±1.68 |
| Dataset | Cross-scenario/scan real-time imaging | | | | |
| | 100% | 75% | 50% | 25% | 20% |
| Physical phantom | 1.45±1.20 | 1.66±1.43 | 1.91±1.78 | 1.94±1.76 | 2.52±2.04 |
| Clinical | 1.11±0.98 | 1.47±0.99 | 1.78±1.45 | 1.96±1.28 | 2.75±2.27 |

**5. Discussion**

In this study, we introduced DREME-GSMR, an innovative framework for dynamic volumetric MRI reconstruction and real-time imaging based on Gaussian representations. DREME-GSMR simultaneously reconstructs time-resolved dynamic MRIs and learns a motion model capable of resolving inter-frame motion directly from pre-treatment k-space data in a 'one-shot' fashion, without requiring predefined anatomical or motion models. The motion model can be subsequently applied to future intra-treatment k-space data, for real-time imaging and motion estimation.

In DREME-GSMR, we designed a dual-path motion encoder to combine the complementary strengths of MLP-based and CNN-based motion encoders. The MLP branch uses minimal k-space center information to predict temporal motion scores and enables on-the-fly temporal binning during the early training stages, which stabilizes optimization and facilitates coarse-to-fine learning. In contrast, the CNN branch processes 2D FFT projection image inputs and supports motion-augmentation training using simulated data with unseen motion patterns. Our ablation study showed that, under a similar augmentation scheme, the MLP-only variant was less effective than the dual-path model for real-time motion estimation. A likely reason is that the NUFFT-simulated k-space signals used for augmentation are based on very limited k-space data and do not fully reproduce the coil-sensitivity characteristics and noise properties of the acquired data. By comparison, the CNN branch can extract motion information from the visual structure of the simulated projections and is therefore more robust to this simulation-to-measurement mismatch. **Figure 12** provides a representative example of the effectiveness of motion augmentation training. When the model was trained on a scan with a smaller motion range and tested on a scan with a substantially larger unseen motion range, motion-augmentation training enabled DREME-GSMR to better recover the target trajectory, whereas the model trained without augmentation showed more pronounced motion undershooting. These findings support the role of the CNN branch and motion-augmentation strategy in improving generalizability to unseen motion patterns.

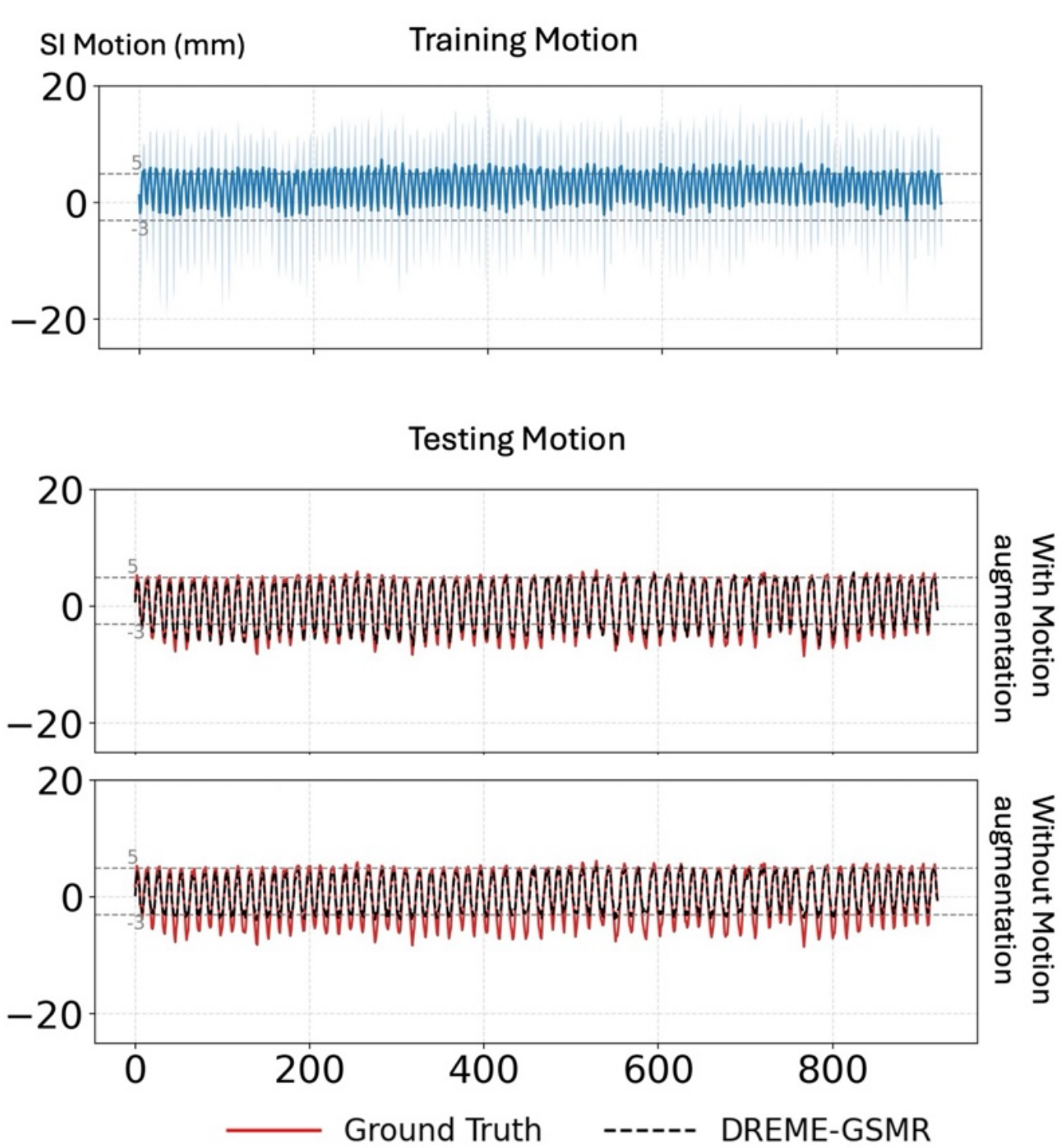

**Figure 12.** An example illustrating the effect of motion-augmentation training on cross-scan real-time motion estimation. The top row shows the training motion from volunteer V4, scan B, which has a peak-to-valley motion range of 4.06 ± 0.78 mm. The blue band shows the motion augmentation range during training. The middle and bottom rows show testing on scan A of the same case, which has a motion range of 10.87 ± 1.05 mm, with and without motion augmentation, respectively. The model trained with motion augmentation shows closer agreement with the ground-truth trajectory and better recovery of the larger unseen motion amplitude.

To evaluate real-time performance on clinical data, we adopted a cross-scan validation strategy, because real-time high-resolution 3D images cannot be acquired in-vivo and no absolute ground truth is available. For each subject, two scans (A and B) were acquired. A model trained on scan A performed real-time inference on the online signal of scan B, and the output was compared against the dynamic reconstruction of a reference model trained on scan B (and vice versa). This design mirrors the intended clinical workflow, in which a pre-treatment scan is available for model training while real-time inference must be performed online on a separately acquired scan. The resulting COME quantifies agreement with a model-derived reference rather than absolute tracking accuracy. This model-derived reference was generated using the same procedure that has been validated against known ground truth in the digital and physical phantom studies, where DREME-GSMR recovered the true dynamic volumes and motion range to around 1 mm. The phantom experiments therefore establish the accuracy and reconstruction fidelity of the framework and show the deviation of the reference from true motion. A prior real-time volumetric study, MRSIGMA (Feng *et al.*, 2020), used real-time 2D x-z projections as a model-independent reference for motion tracking. At rotation angles where the liver dome was visible, an x-z projection was reconstructed directly from the acquired radial-stack k-space, and the SI distance from the liver dome to the FOV edge was compared with that measured on the matched slice of the real-time 3D image. Although this reference is directly data-derived, it does not produce a continuous motion trace, requires manual measurement of the liver dome, and is limited to the SI displacement of a single landmark, effectively treating the liver as rigid. Our model-derived reference provides a dense, temporally continuous, and deformable comparison. We therefore consider the cross-scan strategy better suited to evaluating the proposed framework, with the understanding that the clinical metrics reported here reflect agreement with the 'pseudo-ground truth' and should be interpreted alongside the phantom results for absolute accuracy.

In the clinical workflow, after the patient is set up, a pre-treatment MRI is acquired and used to optimize DREME-GSMR, which reconstructs the dynamic MRI and yields a patient-specific motion model that can subsequently produce real-time volumetric images during beam delivery. For MRI-guided radiotherapy, the end-to-end latency of real-time adaptation should remain below 500 ms (Keall *et al.*, 2021b) for respiratory motion. The current DREME-GSMR implementation reconstructs volumes at a temporal resolution of ~400 ms with an inference time of ~10 ms per volume. Therefore, once the motion model is trained, inference is not the limiting factor. The optimization stage of DREME-GSMR, however, currently takes ~50 min on an NVIDIA RTX 4090 GPU, which is the main limitation. In the future, the training time of DREME-GSMR can be further reduced by employing more advanced hardware (faster GPU cards, for instance, RTX 5090). Additionally, accelerated 3DGS frameworks like FlashGS (Feng *et al.*, 2024) can be integrated into our Gaussian framework to speed up reconstruction fundamentally. Patient-specific priors can also be exploited to accelerate training. Our recently introduced DREME-adapt framework (Zuo *et al.*, 2025) performs a one-time 'virtual-fraction' reconstruction from the pre-treatment 4D-CT, then warm-starts subsequent fractions with the reference CBCT and motion model solved by the preceding fraction in a daisy chain fashion, which has demonstrated an 85% reduction in training time while maintaining accuracy. Practically, with advanced hardware, using DREME-adapt for DREME-GSMR could cut the reconstruction time from ~50 minutes down to ~5 minutes, which would potentially meet clinical timing requirements.

## 5. Conclusion

DREME-GSMR provides a spatiotemporal Gaussian representation framework for patient-specific dynamic MRI reconstruction and real-time motion inference from stack-of-stars k-space data. By jointly learning a reference-frame MRI, motion-basis components, and a dual-path motion encoder, the framework enables dynamic reconstruction from a pre-treatment acquisition and subsequent real-time volumetric inference from online MR signals. Validation on XCAT simulations, physical phantom measurements, and MR-LINAC volunteer and patient datasets demonstrated accurate motion tracking, improved reconstruction quality, and reduced training time compared with DREME-MR. Overall, DREME-GSMR advances the clinical feasibility of time-resolved volumetric MRI for motion-adaptive radiotherapy and other motion-sensitive treatments.

**Funding sources**

The study was supported by the US National Institutes of Health (R01 CA240808, R01 CA258987, R01 EB034691, and R01 CA280135).

**Declaration of competing interests**

The authors declare that they have no known competing financial interests or personal relationships that could have appeared to influence the work reported in this paper.

**Data availability**

The UTSW patient dataset was retrospectively collected in 2025 and 2026 as part of an approved study at UT Southwestern Medical Center, under an umbrella IRB protocol 082013-008 (Improving radiation treatment quality and safety by retrospective data analysis). The volunteer dataset was acquired at UTSW by a separate IRB protocol (STU2021-0206). This is an analysis study and not a clinical trial. Individual volunteer/patient consent was signed for the anonymized use of the imaging data for analysis. Our in-house dataset is not publicly available because it contains sensitive and confidential patient information.

**References**

Ahmed A H, Zhou R, Yang Y, Nagpal P, Salerno M and Jacob M 2020 Free-Breathing and Ungated Dynamic MRI Using Navigator-Less Spiral SToRM *IEEE Trans Med Imaging* **39** 3933-43

Ball H J, Santanam L, Senan S, Tanyi J A, van Herk M and Keall P J 2022 Results from the AAPM Task Group 324 respiratory motion management in radiation oncology survey *Journal of applied clinical medical physics* **23** e13810

Bertholet J, Knopf A, Eiben B, McClelland J, Grimwood A, Harris E, Menten M, Poulsen P, Nguyen D T and Keall P 2019 Real-time intrafraction motion monitoring in external beam radiotherapy *Physics in medicine & biology* **64** 15TR01

Bertholet J, Worm E S, Fledelius W, Høyer M and Poulsen P R 2016 Time-resolved intrafraction target translations and rotations during stereotactic liver radiation therapy: implications for marker-based localization accuracy *International Journal of Radiation Oncology* Biology* Physics* **95** 802-9

Chen J, Huang C, Shanbhogue K, Xia D, Bruno M, Huang Y, Block K T, Chandarana H and Feng L 2024a DCE-MRI of the liver with sub-second temporal resolution using GRASP-Pro with navi-stack-of-stars sampling *NMR Biomed* **37** e5262

Chen J, Liu Y, Wei S, Bian Z, Subramanian S, Carass A, Prince J L and Du Y 2025 A survey on deep learning in medical image registration: New technologies, uncertainty, evaluation metrics, and beyond *Med Image Anal* **100** 103385

Chen J, Xia D, Huang C, Shanbhogue K, Chandarana H and Feng L 2024b Free-breathing time-resolved 4D MRI with improved T1-weighting contrast *NMR Biomed* **37** e5247

Chun S Y and Fessler J A 2009 A simple regularizer for B-spline nonrigid image registration that encourages local invertibility *IEEE J Sel Top Signal Process* **3** 159-69

Corradini S, Alongi F, Andratschke N, Belka C, Boldrini L, Cellini F, Debus J, Guckenberger M, Hörner-Rieber J and Lagerwaard F 2019 MR-guidance in clinical reality: current treatment challenges and future perspectives *Radiation Oncology* **14** 1-12

Daly M, McDaid L, Anandadas C, Brocklehurst A, Choudhury A, McWilliam A, Radhakrishna G and Eccles C L 2024 Impact of motion management strategies on abdominal organ at risk delineation for magnetic resonance-guided radiotherapy *Phys Imaging Radiat Oncol* **32** 100650

Dietrich O, Raya J G, Reeder S B, Reiser M F and Schoenberg S O 2007 Measurement of signal-to-noise ratios in MR images: influence of multichannel coils, parallel imaging, and reconstruction filters *J Magn Reson Imaging* **26** 375-85

Fei B, Xu J, Zhang R, Zhou Q, Yang W and He Y 2024 3D Gaussian Splatting as New Era: A Survey *IEEE transactions on visualization and computer graphics* **PP**

Feng G, Chen S, Fu R, Liao Z, Wang Y, Liu T, Pei Z, Li H, Zhang X and Dai B 2024 Flashgs: Efficient 3d gaussian splatting for large-scale and high-resolution rendering *arXiv preprint arXiv:2408.07967*

Feng L 2022 Golden‐angle radial MRI: basics, advances, and applications *Journal of Magnetic Resonance Imaging* **56** 45-62

Feng L 2023a 4D Golden-Angle Radial MRI at Subsecond Temporal Resolution *NMR Biomed* **36** e4844

Feng L 2023b Live-view 4D GRASP MRI: A framework for robust real-time respiratory motion tracking with a sub-second imaging latency *Magn Reson Med* **90** 1053-68

Feng L, Axel L, Chandarana H, Block K T, Sodickson D K and Otazo R 2016 XD-GRASP: Golden-angle radial MRI with reconstruction of extra motion-state dimensions using compressed sensing *Magn Reson Med* **75** 775-88

Feng L, Tyagi N and Otazo R 2020 MRSIGMA: Magnetic Resonance SIGnature MAtching for real-time volumetric imaging *Magn Reson Med* **84** 1280-92

Ferris W S, George B, Plichta K A, Caster J M, Hyer D E, Smith B R and St-Aubin J J 2024 Clinical experience with adaptive MRI-guided pancreatic SBRT and the use of abdominal compression to reduce treatment volume *Front Oncol* **14** 1441227

Gudbjartsson H and Patz S 1995 The Rician distribution of noisy MRI data *Magn Reson Med* **34** 910-4

Hall W A, Paulson E S, van der Heide U A, Fuller C D, Raaymakers B, Lagendijk J J, Li X A, Jaffray D A, Dawson L A and Erickson B 2019 The transformation of radiation oncology using real-time magnetic resonance guidance: A review *European Journal of Cancer* **122** 42-52

Heyde B, Alessandrini M, Hermans J, Barbosa D, Claus P and D'Hooge J 2016 Anatomical Image Registration Using Volume Conservation to Assess Cardiac Deformation From 3D Ultrasound Recordings *IEEE Trans Med Imaging* **35** 501-11

Huttinga N R, Bruijnen T, van den Berg C A and Sbrizzi A 2021a Nonrigid 3D motion estimation at high temporal resolution from prospectively undersampled k‐space data using low‐rank MR‐MOTUS *Magnetic resonance in medicine* **85** 2309-26

Huttinga N R, Van den Berg C A, Luijten P R and Sbrizzi A 2020 MR-MOTUS: model-based non-rigid motion estimation for MR-guided radiotherapy using a reference image and minimal k-space data *Physics in Medicine & Biology* **65** 015004

Huttinga N R F, Bruijnen T, van den Berg C A T and Sbrizzi A 2021b Nonrigid 3D motion estimation at high temporal resolution from prospectively undersampled k-space data using low-rank MR-MOTUS *Magn Reson Med* **85** 2309-26

Huttinga N R F, Bruijnen T, Van Den Berg C A T and Sbrizzi A 2022 Real-Time Non-Rigid 3D Respiratory Motion Estimation for MR-Guided Radiotherapy Using MR-MOTUS *IEEE Trans Med Imaging* **41** 332-46

Jiang W, Ong F, Johnson K M, Nagle S K, Hope T A, Lustig M and Larson P E Z 2018 Motion robust high resolution 3D free-breathing pulmonary MRI using dynamic 3D image self-navigator *Magn Reson Med* **79** 2954-67

Kaufman L, Kramer D M, Crooks L E and Ortendahl D A 1989 Measuring signal-to-noise ratios in MR imaging *Radiology* **173** 265-7

Keall P, Poulsen P and Booth J T *Seminars in radiation oncology,2019),* vol. Series 29*)*: Elsevier) pp 228-35

Keall P J, El Naqa I, Fast M F, Hewson E A, Hindley N, Poulsen P, Sengupta C, Tyagi N and Waddington D E 2025 Real-time dose-guided radiation therapy *International Journal of Radiation Oncology* Biology* Physics* **122** 787-801

Keall P J, Sawant A, Berbeco R I, Booth J T, Cho B, Cerviño L I, Cirino E, Dieterich S, Fast M F and Greer P B 2021a AAPM Task Group 264: The safe clinical implementation of MLC tracking in radiotherapy *Medical physics* **48** e44-e64

Keall P J, Sawant A, Berbeco R I, Booth J T, Cho B, Cerviño L I, Cirino E, Dieterich S, Fast M F, Greer P B, Munck af Rosenschöld P, Parikh P J, Poulsen P R, Santanam L, Sherouse G W, Shi J and Stathakis S 2021b AAPM Task Group 264: The safe clinical implementation of MLC tracking in radiotherapy *Medical Physics* **48** e44-e64

Keesman R, van der Bijl E, Kerkmeijer L G W, Tyagi N, Akdag O, Wolthaus J W H, van de Pol S M G, Noteboom J L, Intven M P W, Fast M F and van Lier A 2024 Multi-institutional experience treating patients with cardiac devices on a 1.5 Tesla magnetic resonance-linear accelerator and workflow development for thoracic treatments *Phys Imaging Radiat Oncol* **32** 100680

Kingma D P and Ba J 2014 Adam: A Method for Stochastic Optimization *CoRR* **abs/1412.6980**

Li R, Lewis J H, Jia X, Zhao T, Liu W, Wuenschel S, Lamb J, Yang D, Low D A and Jiang S B 2011 On a PCA-based lung motion model *Physics in Medicine & Biology* **56** 6009

Li Y, Fu X, Li H, Zhao S, Jin R and Zhou S K 2025 3DGR-CT: Sparse-view CT reconstruction with a 3D Gaussian representation *Medical Image Analysis* 103585

Li Z, Huang C, Tong A, Chandarana H and Feng L 2023 Kz-accelerated variable-density stack-of-stars MRI *Magn Reson Imaging* **97** 56-67

Lin Y, Wang H, Chen J and Li X *International Conference on Medical Image Computing and Computer-Assisted Intervention,2024),* vol. Series*)*: Springer) pp 425-35

Lombardo E, Dhont J, Page D, Garibaldi C, Künzel L A, Hurkmans C, Tijssen R H, Paganelli C, Liu P Z and Keall P J 2024 Real-time motion management in MRI-guided radiotherapy: Current status and AI-enabled prospects *Radiotherapy and Oncology* **190** 109970

McNair H and Buijs M 2019 Image guided radiotherapy moving towards real time adaptive radiotherapy; global positioning system for radiotherapy? *Technical Innovations & Patient Support in Radiation Oncology* **12** 1

Menchón-Lara R-M, Simmross-Wattenberg F, Casaseca-de-la-Higuera P, Martín-Fernández M and Alberola-López C 2019 Reconstruction techniques for cardiac cine MRI *Insights into imaging* **10** 1-16

Mildenhall B, Srinivasan P P, Tancik M, Barron J T, Ramamoorthi R and Ng R 2021 NeRF: representing scenes as neural radiance fields for view synthesis *Commun. ACM* **65** 99–106

Muckley M J, Stern R, Murrell T and Knoll F*,* vol. Series*)*

Nayak K S 2019 Response to Letter to the Editor:"Nomenclature for real - time magnetic resonance imaging" *Magnetic resonance in medicine* **82**

Nayak K S, Lim Y, Campbell - Washburn A E and Steeden J 2022 Real - time magnetic resonance imaging *Journal of Magnetic Resonance Imaging* **55** 81-99

Ong F, Zhu X, Cheng J Y, Johnson K M, Larson P E, Vasanawala S S and Lustig M 2020 Extreme MRI: Large - scale volumetric dynamic imaging from continuous non - gated acquisitions *Magnetic resonance in medicine* **84** 1763-80

Owrangi A M, Greer P B and Glide-Hurst C K 2018 MRI-only treatment planning: benefits and challenges *Physics in Medicine & Biology* **63** 05TR1

Paganelli C, Whelan B, Peroni M, Summers P, Fast M, van de Lindt T, McClelland J, Eiben B, Keall P and Lomax T 2018 MRI-guidance for motion management in external beam radiotherapy: current status and future challenges *Physics in Medicine & Biology* **63** 22TR03

Paszke A, Gross S, Massa F, Lerer A, Bradbury J, Chanan G, Killeen T, Lin Z, Gimelshein N, Antiga L, Desmaison A, Köpf A, Yang E, DeVito Z, Raison M, Tejani A, Chilamkurthy S, Steiner B, Fang L, Bai J and Chintala S 2019 PyTorch: An Imperative Style, High-Performance Deep Learning Library *ArXiv* **abs/1912.01703**

Paulson E S, Ahunbay E, Chen X, Mickevicius N J, Chen G P, Schultz C, Erickson B, Straza M, Hall W A and Li X A 2020 4D-MRI driven MR-guided online adaptive radiotherapy for abdominal stereotactic body radiation therapy on a high field MR-Linac: Implementation and initial clinical experience *Clin Transl Radiat Oncol* **23** 72-9

Peng T, Zha R, Li Z, Liu X and Zou Q 2025a Three-Dimensional MRI Reconstruction with Gaussian Representations: Tackling the Undersampling Problem *arXiv preprint arXiv:2502.06510*

Peng T, Zha R and Zou Q 2025b *MoRe-3DGSMR: Motion-resolved reconstruction framework for free-breathing pulmonary MRI based on 3D Gaussian representation*

Rajiah P S, François C J and Leiner T 2023 Cardiac MRI: state of the art *Radiology* **307** e223008

Segars W P, Sturgeon G, Mendonca S, Grimes J and Tsui B M 2010 4D XCAT phantom for multimodality imaging research *Med Phys* **37** 4902-15

Seppenwoolde Y, Shirato H, Kitamura K, Shimizu S, Van Herk M, Lebesque J V and Miyasaka K 2002 Precise and real-time measurement of 3D tumor motion in lung due to breathing and heartbeat, measured during radiotherapy *International Journal of Radiation Oncology* Biology* Physics* **53** 822-34

Shao H-C, Li T, Dohopolski M J, Wang J, Cai J, Tan J, Wang K and Zhang Y 2022 Real-time MRI motion estimation through an unsupervised k-space-driven deformable registration network (KS-RegNet) *Physics in Medicine & Biology* **67** 135012

Shao H-C, Mengke T, Deng J and Zhang Y 2024a 3D cine-magnetic resonance imaging using spatial and temporal implicit neural representation learning (STINR-MR) *Physics in Medicine & Biology* **69** 095007

Shao H-C, Mengke T, Pan T and Zhang Y 2024b Dynamic CBCT imaging using prior model-free spatiotemporal implicit neural representation (PMF-STINR) *Physics in Medicine & Biology* **69** 115030

Shao H-C, Mengke T, Pan T and Zhang Y 2025a Real-time CBCT imaging and motion tracking via a single arbitrarily-angled x-ray projection by a joint dynamic reconstruction and motion estimation (DREME) framework *Physics in Medicine & Biology* **70** 025026

Shao H-C, Qian X, Xu G, Wu C, Otazo R, Deng J and Zhang Y 2025b A dynamic reconstruction and motion estimation framework for cardiorespiratory motion-resolved real-time volumetric MR imaging (DREME-MR) *Physics in Medicine & Biology* **70** 175013

Shen L, Pauly J and Xing L 2022 NeRP: Implicit Neural Representation Learning With Prior Embedding for Sparsely Sampled Image Reconstruction *IEEE Trans Neural Netw Learn Syst* **Pp**

Siddiq S, Murray V, Tyagi N, Borman P, Gui C, Crane C, Wu C and Otazo R 2024 MR signature matching (MRSIGMA) implementation for true real-time free-breathing volumetric imaging with sub-200 ms latency on an MR-Linac *Magnetic Resonance in Medicine* **92** 1162-76

Stemkens B, Paulson E S and Tijssen R H 2018 Nuts and bolts of 4D-MRI for radiotherapy *Physics in Medicine & Biology* **63** 21TR01

Stemkens B, Tijssen R H, De Senneville B D, Lagendijk J J and Van Den Berg C A 2016 Image-driven, model-based 3D abdominal motion estimation for MR-guided radiotherapy *Physics in Medicine & Biology* **61** 5335

Terpstra M L, Maspero M, Bruijnen T, Verhoeff J J C, Lagendijk J J W and van den Berg C A T 2021 Real-time 3D motion estimation from undersampled MRI using multi-resolution neural networks *Medical Physics* **48** 6597-613

Wei R, Chen J, Liang B, Chen X, Men K and Dai J 2023 Real-time 3D MRI reconstruction from cine-MRI using unsupervised network in MRI-guided radiotherapy for liver cancer *Medical Physics* **50** 3584-96

Wu S, Guo Y, Ji Y, Tong J, Lu Y, Li M, Huang S, Ding Y and Lu H 2024 GBIR: A Novel Gaussian Iterative Method for Medical Image Reconstruction

Xie J, Shao H C and Zhang Y 2025 Time-resolved dynamic CBCT reconstruction using prior-model-free spatiotemporal Gaussian representation (PMF-STGR) *Phys Med Biol* **70**

Yu S, Dai G, Wang Z, Li L, Wei X and Xie Y 2018 A consistency evaluation of signal-to-noise ratio in the quality assessment of human brain magnetic resonance images *BMC Med Imaging* **18** 17

Zha R, Lin T J, Cai Y, Cao J, Zhang Y and Li H 2024 R $^ 2$-Gaussian: Rectifying Radiative Gaussian Splatting for Tomographic Reconstruction *arXiv preprint arXiv:2405.20693*

Zhang Q, Pevsner A, Hertanto A, Hu Y C, Rosenzweig K E, Ling C C and Mageras G S 2007 A patient - specific respiratory model of anatomical motion for radiation treatment planning *Medical physics* **34** 4772-81

Zhang X, Xie G, Lu N, Zhu Y, Wei Z, Su S, Shi C, Yan F, Liu X, Qiu B and Fan Z 2019 3D self-gated cardiac cine imaging at 3 Tesla using stack-of-stars bSSFP with tiny golden angles and compressed sensing *Magn Reson Med* **81** 3234-44

Zhang Y, Ma J, Iyengar P, Zhong Y and Wang J 2017 A new CT reconstruction technique using adaptive deformation recovery and intensity correction (ADRIC) *Med Phys* **44** 2223-41

Zhou W, Bovik A C, Sheikh H R and Simoncelli E P 2004 Image quality assessment: from error visibility to structural similarity *IEEE Transactions on Image Processing* **13** 600-12

Zuo R, Shao H-C and Zhang Y 2025 Prior-Adapted Progressive Time-Resolved CBCT Reconstruction Using a Dynamic Reconstruction and Motion Estimation Method *arXiv preprint arXiv:2504.18700*